\documentclass[onecolumn,preprintnumbers,amsmath,amssymb]{revtex4}

\usepackage{subcaption}
\usepackage{dcolumn}
\usepackage{comment}
\usepackage{forest}
\useforestlibrary{edges}
\usepackage{braket}
\usepackage[normalem]{ulem}
\usepackage{placeins}
\usepackage{float}

\usepackage{hyperref}

\newcommand{\dikicomment}[1]{{\color{blue} \bf [Diki: #1]}}
\newcommand{\rbm}[1]{{\color{red} \bf [Robb: #1]}}
\newcommand{\dy}[1]{{\color{purple} \bf [Dong-han: #1]}}

\newcommand{\Xpolar}{\left|A\ell m\tilde{\Omega}(r,\theta)\sqrt{\left|\frac{P(0)}{f(r)}\right|}\right|}
\newcommand{\Xprolate}{\left|A\ell \Omega(x,y)\sqrt{\left|\frac{P(0)}{P(y)}\right|}\right|}
\newcommand{\bX}{{\bar{X}}}
\newcommand{\ella}{\ell_{\alpha}}
\DeclareMathOperator{\sgn}{\text{sgn}}

\usepackage{amsmath,latexsym,amssymb,amsfonts}

\begin{document}

\title{\textbf{Acceleration in 3D Einstein-Gauss-Bonnet Gravity}}

\author{
Cendikiawan Suryaatmadja$^{a}$\footnote{{\tt cendikiawan.suryaatmadja@uwaterloo.ca}},
Robert B. Mann$^{a,b}$\footnote{{\tt rbmann@uwaterloo.ca}}
and
Dong-han Yeom$^{a,b,c,d}$\footnote{{\tt innocent.yeom@gmail.com}}
}

\affiliation{
$^{a}$Department of Physics and Astronomy, University of Waterloo, Waterloo, ON N2L 3G1, Canada\\
$^{b}$Perimeter Institute, Waterloo, ON N2L 2Y5, Canada\\
$^{c}$Department of Physics Education, Pusan National University, Busan 46241, Republic of Korea\\
$^{d}$Leung Center for Cosmology and Particle Astrophysics, National Taiwan University, Taipei 10617, Taiwan
}

\begin{abstract}
    We present a new class of exact solutions in  Einstein-Gauss-Bonnet gravity in 2+1 dimensions that generalize the C-metric. This set of metrics equals the C-metric multiplied by a factor which, along with the massless scalar field, depends on a single variable whose value governs the structure of the spacetime. As in Einstein gravity there are three classes of metrics, but within each class
    we find six distinct subclasses of solutions. After discussing their basic structure, we  concentrate only on one subclass that is locally AdS.
    In the zero-coupling limit, this subclass of solutions not only remains well defined and recovers the C-metric but also encompasses two previously unknown representations of the AdS spacetime. Furthermore, we establish the existence of a domain wall and delineate its energy conditions.   We also find new classes of solutions of non-constant curvature, whose interpretation remains to be understood.  
\end{abstract}

\maketitle

\newpage

\tableofcontents


\newpage

\section{Introduction}

One of the most intriguing solutions to general relativity is the C-metric \cite{article,weyl1919statischen}. For many years, its physical relevance remained unclear until it was shown that it can be understood as a black hole undergoing uniform acceleration \cite{kinnersley1970uniformly}.  A topological defect of dimension one (a string) provides an acceleration force with either positive or negative tension. This metric has been extensively studied in recent years, with generalizations to rotating and charged black holes, and to the inclusion of a cosmological constant $\Lambda$ \cite{Plebanski:1976gy,Mann:1995vb,Podolsky:2002,Podolsk__2006,Dias_2003}. Typically, such solutions contain a second black hole, isolated from its partner by a non-compact acceleration horizon. However, if $\Lambda < 0$, there is a class of `slowly' accelerating black holes that exhibit 
neither an acceleration horizon nor a second black hole. Rather, the spacetime consists of a single black hole suspended at a fixed distance from the center of the anti-de Sitter (AdS), i.e., in a locally accelerating frame. This class of black holes has yielded important advances in our understanding of the role of acceleration. After some effort \cite{Astorino:2016xiy,Astorino:2016ybm,Appels:2016uha},
a consistent thermodynamics has been established for these objects
\cite{Appels:2017xoe,Anabalon:2018qfv,Gregory:2019dtq}, with extensions to `rapidly' accelerating black holes that have an acceleration horizon \cite{Gregory:2020mmi}. 

However, such generalizations have not included higher-curvature effects. Terms non-linear in the curvature are expected to modify general relativity due to quantum effects, with string theory providing numerous examples.  The simplest such generalization is Einstein-Gauss-Bonnet gravity. The action of this theory is as follows: 
\begin{eqnarray}\label{EinHilG}
    S = \int d^D x\sqrt{-g} \left[ R-2\Lambda+\alpha \mathcal{G}  \right],
\end{eqnarray}
where $R$ is the Ricci scalar, and the additional term
\begin{eqnarray}
    \mathcal{G} \equiv R^2-4R^{\mu\nu}R_{\mu\nu}+R^{\mu\nu\rho\sigma}R_{\mu\nu\rho\sigma}
\end{eqnarray}
is known as the Gauss-Bonnet term. Here, $\alpha>0$ represents the coupling constant. Recently, an accelerating black hole solution to the field equations of Eq.~\eqref{EinHilG} for $D=5$ dimensions was obtained for a particular choice of the coupling $\alpha$ (known as the Chern-Simons point) by embedding a 3-dimensional accelerating solution into a larger space of hyperbolic curvature \cite{Anabalon:2024abz}. 

Einstein-Gauss-Bonnet gravity becomes trivial in (3+1) dimensions since the additional term $\mathcal{G}$ is a topological invariant in less than $D<5$ spacetime dimensions. However, it is
possible to obtain a 
$D\to 4$ limit of this theory \cite{Hennigar_2020,Lu:2020iav,Fernandes:2020nbq}.  Known as 4-dimensional Einstein-Gauss-Bonnet gravity (4DEGB), it was constructed from 
the observation that solutions to the $n$-dimensional field equations of Eq.~\eqref{EinHilG} have a well-defined $D\to 4$ limit provided $\alpha$ was appropriately rescaled \cite{Glavan:2019inb}. The action is given by \cite{Hennigar_2020,Fernandes:2020nbq,Lu:2020iav}:
\begin{eqnarray}
    S = \int d^4 x\sqrt{-g} \left[ R-2\Lambda+\alpha \mathcal{L}_{\phi} \right],
\label{eq1}
\end{eqnarray}
  where 
\begin{eqnarray}
    \mathcal{L}_{\phi} \equiv \phi\mathcal{G}+4 G^{ab}\partial_a\phi\partial_b\phi-4 (\partial\phi)^2\Box\phi+2(\partial \phi)^4,\label{Lphi}
\end{eqnarray}
and includes the scalar field $\phi$. This theory has a number of interesting phenomenological consequences and has been the subject of much investigation \cite{Fernandes_2022}.

Of course, once constructed, it is possible to consider the action, Eq.~\eqref{eq1}, in any spacetime dimension. The $D=3$ case -- 3DEGB --  has been of particular interest, and a number of exact solutions have been found \cite{Hennigar:2020fkv}. These solutions generalize the Banados-Teitelboim-Zanelli (BTZ) black hole solution in $D=3$ general relativity (with $\alpha=0$) \cite{Ba_ados_1992}.
3DEGB has been extensively investigated, with many interesting solutions and features discovered \cite{Konoplya:2020ibi,Alkac:2020zhg,Lu:2020mjp,Ma:2020ufk,Hennigar:2020drx, Narzilloev:2021jtg,Dimov:2021fbm,Cuadros-Melgar:2022lrf,Ahmed:2022dpu,Jusufi:2023fpo,Alkac:2023mvr,Skvortsova:2023zca,Alkac:2024hvu}.

Motivated by the above, we focus in this paper on 3-dimensional Einstein-Gauss-Bonnet gravity (3DEGB) to obtain generalizations of the   C-metric.
 Our solutions 
are higher-curvature generalizations of those
 previously obtained in Einstein gravity in (2+1) dimensions \cite{Astorino_2011}, which included black-hole and point-mass solutions \cite{ArenasHenriquez:2023hur,Bunney:2024xic}. Accelerating hairy black-hole solutions in $(2+1)$ dimensions with a conformally coupled scalar field were constructed in \cite{Cisterna_2023}.
These C-metric generalizations generically do not have circular symmetry. 
They are regular everywhere, except for a discontinuity in the curvatures at one angular direction; one needs to introduce a thin string (technically, a domain wall) to satisfy the junction condition.

To obtain accelerating solutions to 3DEGB, we introduce a new metric ansatz, which we describe below.  We thereby obtain a new set of C-metric-like solutions.
As in Einstein gravity \cite{Arenas_Henriquez_2022}
we obtain  three classes of metrics, but within each class
    we find six distinct subclasses of solutions. 
Like the original C-metric, our solutions require a thin string (technically a domain wall in $D=3$) at one angular direction to satisfy the junction conditions. It is interesting to note that in the Einstein limit, we can reproduce C-metric solutions; however, some solution branches do not reduce to C-metric solutions. These represent new classes of solutions even in the pure Einstein gravity limit.

Our paper is organized as follows. In Sec.~\ref{sec:3d}, we write the equations of motion of the 3D Einstein-Gauss-Bonnet gravity; we also summarize the three previously known classes of C-metric solutions \cite{Arenas_Henriquez_2022} in   Einstein gravity.   In Sec.~\ref{sec:cla}, we introduce a new metric ansatz that preserves these classes, 
and use it to
find six distinct subclasses of solutions within each class. In Sec.~\ref{sec:ana}, we focus in particular on the details of one subclass. In Sec.~\ref{sec:polar}, we discuss the polar coordinate representation and its domain conventions; based on this, in Sec.~\ref{sec:spa}, we present analytical and numerical details of the spacetime in both prolate and polar coordinates. In Sec.~\ref{sec:pro}, we study the physical properties of thin strings along the curvature discontinuities. Finally, in Sec.~\ref{sec:dis}, we summarize our results and discuss the physical meaning of our new solutions. In this paper, we use the conventions $ c = G = \hbar = 1$ and $\Lambda=-1/\ell^2$. 


\section{\label{sec:3d}Preliminaries}

Before we focus on new solutions, we summarize the basic features of 3DEGB.  First, we will discuss the equations of motion of 3DEGB. We then review the C-metric solutions in Einstein gravity  \cite{Astorino_2011,Arenas_Henriquez_2022,ArenasHenriquez:2023hur,Bunney:2024xic} to set the stage for the new solutions that we obtain.


\subsection{2+1D Einstein-Gauss-Bonnet gravity}

Variation of the action, Eq.~\eqref{eq1}, with respect to the metric in $D$-dimensions yields the field equations $\mathcal{E}_{ab}=0$, where
\begin{eqnarray}
    \mathcal{E}_{ab} &=& \Lambda g_{ab} + G_{ab} \nonumber\\
    && + \alpha \left(\phi H_{ab} - 2R \left( \nabla_{a} \phi \nabla_{b} \phi + \nabla_{b} \nabla_{a} \phi \right) + 8R^c_{\,(a} \nabla_{b)}\nabla_{c} \phi+ 8R^c_{(a} \nabla_{b)} \phi\nabla_{c} \right. \phi\nonumber\\
    &&- 2G_{ab} \left( (\nabla \phi)^{2} + 2 \Box \phi \right) - 4 \left( \nabla_{a} \phi\nabla_{b} \phi + \nabla_{b} \nabla_{a} \phi \right) \Box \phi - \left( g_{ab} (\nabla \phi)^{2} -4 \nabla_a\phi \nabla_b\phi \right) (\nabla \phi)^{2} \nonumber \\
    &&+ 8 \nabla_{(a} \phi \nabla_{b)} \nabla_{c} \phi  \nabla^{c} \phi - 4 g_{ab} R^{cd} \left( \nabla_{c} \nabla_{d} \phi + \nabla_{c} \phi \nabla_{d} \phi \right) + 2g_{ab} (\Box \phi)^{2} - 2g_{ab} \nabla_{c} \nabla_{d} \phi \nabla^{c} \nabla^{d} \phi \nonumber\\
    && \left.- 4 g_{ab} \nabla^{c} \phi \nabla^{d} \phi \nabla_{c} \nabla_{d} \phi+ 4 \nabla_{c} \nabla_{b} \phi \nabla^{c} \nabla_{a} \phi + 4 R_{acbd} \left( \nabla^{c} \phi \nabla^{d} \phi + \nabla^{d} \nabla^{c} \phi \right) \right).
    \label{Eab}
\end{eqnarray}
Here,
\begin{eqnarray} H_{ab} \equiv 2\left(RR_{ab}-2R_{acbd}R^{cd}+R_{acde}R_b{}^{cde}-2R_{ac}R^c{}_b-\frac{1}{4}\mathcal{G}g_{ab}\right),
\end{eqnarray}
with $H_{ab} = 0$ in dimensions $D\leq4$. 

For $D=3$, we can further simplify Eq.~\eqref{Eab} thanks to the identity of the Riemann tensor 
\begin{eqnarray}
    R_{acbd} = G_{ab}g_{cd}-R_{ad}g_{bc}+R_{cd}g_{ab}-R_{bc}g_{ad}+\frac{1}{2}Rg_{ad}g_{bc}
\end{eqnarray}
valid in (2+1) dimensions. 
From this, one can derive 
\begin{eqnarray}
    4R_{acbd}\left( \nabla^c\phi \nabla^d\phi+\nabla^d\nabla^c\phi \right) &=& 4G_{ab}\left( (\nabla\phi)^2+\Box\phi \right)
    -4R_{ad}g_{bc} \left( \nabla^c\phi \nabla^d\phi + \nabla^d\nabla^c\phi \right) \nonumber \\
    && +4g_{ab}R^{cd} \left(\nabla_c\nabla_d\phi+\nabla_c\phi \nabla_d\phi \right) -4R_{bc}g_{ad} \left( \nabla^c\phi \nabla^d\phi+\nabla^d\nabla^c\phi \right) \nonumber \\
   && +2R \left( \nabla_a\phi \nabla_b\phi+\nabla_a\nabla_b\phi \right)\label{Riemannphiphi}.
\end{eqnarray}
Inserting Eq.~\eqref{Riemannphiphi} into the original equation of motion, Eq.~\eqref{Eab}, we obtain the simpler expression 
\begin{eqnarray}
    \mathcal{E}_{ab}&=& \Lambda g_{ab} + G_{ab}(1+ 2\alpha (\nabla \phi)^{2}) \nonumber \\
    && + \alpha \left(- 4 \left( \nabla_{a} \phi\nabla_{b} \phi + \nabla_{b} \nabla_{a} \phi \right) \Box \phi + 8 \nabla_{(a} \phi \nabla_{b)} \nabla_{c} \phi \nabla^{c} \phi+4 \nabla_a\phi \nabla_b\phi \left(\nabla \phi \right)^{2} + 4 \nabla_{c} \nabla_{b} \phi \nabla^{c} \nabla_{a} \phi \right. \nonumber \\
    && \left.  \qquad\qquad + g_{ab} \left(2 \left(\Box \phi\right)^{2} -  \left(\nabla \phi\right)^4-2 \nabla_{c} \nabla_{d} \phi\nabla^{c} \nabla^{d} \phi
    - 4 \nabla^{c} \phi\nabla^{d} \phi\nabla_{c} \nabla_{d} \phi \right) \right).
    \label{Eabsimp}
\end{eqnarray}

In addition to Eq.~\eqref{Eabsimp}, we also need the equation of motion for $\phi$. Varying the action with respect to $\phi$, we obtain 
\begin{equation}
    \nabla_a\left(-\frac{\partial \mathcal{L}_{\phi}}{\partial (\nabla_a\phi)}+\nabla_b\frac{\partial  \mathcal{L}_{\phi}}{\partial (\nabla_a\nabla_b\phi)}\right) = 0 \label{eq4}
\end{equation}
which implies 
$\mathcal{E}_{\phi}^{a}=$constant, where
\begin{equation}
    \mathcal{E}_\phi^{a} \equiv -\frac{\partial \mathcal{L}_{\phi}}{\partial (\nabla_a\phi)}+\nabla_b\frac{\partial  \mathcal{L}_{\phi}}{\partial (\nabla_a\nabla_b\phi)},\label{zero-field}
\end{equation}
which becomes
\begin{equation}\label{phieq}
   G_{ab} \nabla^b\phi  + \nabla_a\phi\left( 
(\nabla\phi)^2 - \nabla^2\phi\right) +\frac{1}{2}
\nabla_a\left((\nabla\phi)^2 \right) = 0.
\end{equation}
It is straightforward to show that the divergence of this expression yields the standard form \cite{Hennigar:2020fkv} for the field equation for $\phi$. In what follows, to satisfy Eq.~\eqref{eq4}, we will impose the stronger condition  $\mathcal{E}_{\phi}^{a}=0$.

\subsection{C-metric solutions in (2+1)-dimensional Einstein gravity}

Setting $\alpha=0$ in Eq.~\eqref{Eabsimp}, we can obtain a $(2+1)$-dimensional version of the C-metric.  
Written in prolate coordinates $(\tau,y,x)$, all such C-metric solutions conform to the ansatz 
\cite{Astorino_2011,Arenas_Henriquez_2022,ArenasHenriquez:2023hur,Bunney:2024xic}
\begin{eqnarray}\label{cmet}
    g^{C}_{ab}=\frac{1}{\Omega(x,y)^2}\begin{pmatrix}
-P(y)& 0 & 0\\
0 & P(y)^{-1} & 0\\
0 & 0 & Q(x)^{-1}
\end{pmatrix},
\end{eqnarray}
where $\Omega(x,y) \equiv A(x-y)$, and $Q(x)$, $P(y)$ are polynomials, given in Table~\ref{table:: Classes Prolate}, that classify the various solutions. Class~I represents point particles, Class~II represents accelerating black holes, and Class~III represents a nonstandard configuration. The cosmological constant $\Lambda = -1/\ell^2$ and $A$ is the acceleration parameter.  We refer to Class~I as Class~$\mathrm{I}_\mathrm{slow}$ if $A\ell<1$, as Class~$\mathrm{I}_\mathrm{saturated}$ if $A\ell=1$, and as Class~$\mathrm{I}_\mathrm{rapid}$ if $A\ell>1$. 

\begin{table}[H]
\centering
\caption{The three classes of C-metric solutions in prolate coordinates $(x,y)$,  each with a distinct range of $x>y$.}
\begin{tabular}{|c|c|c|c|} 
\hline
 Class & $Q(x)$ & $P(y)$ & Maximal range of $x$ \\ \hline\hline
 I & $1-x^2$ & $\frac{1}{A^2\ell^2}+(y^2-1)$ & $|x|<1$ \\ 
 II & $x^2-1$ & $\frac{1}{A^2\ell^2}+(1-y^2)$ & $x>1$ or $x<-1$ \\ 
 III & $1+x^2$ & $\frac{1}{A^2\ell^2}-(1+y^2)$ & $x\in \mathbb{R}$ \\
 \hline
\end{tabular}
\label{table:: Classes Prolate}
\end{table}

\section{\label{sec:cla} New solutions}

In this section, we will solve the equations $\mathcal{E}_{ab}=0$ and  $\mathcal{E}_{\phi}^{a}=0$ to obtain the metric and the scalar field solutions.
To solve them, we introduce a new metric ansatz in prolate coordinates.

\subsection{Ansatz for new solutions}

We first begin by considering solutions to Eq.~\eqref{phieq}. 
Inserting Eq.~\eqref{cmet} into Eq.~\eqref{phieq}, we obtain the solution
\begin{equation}
    \phi=-\log{X} + \text{const.},
\label{phiC}
\end{equation}
where we define
\begin{eqnarray}\label{Xdef}
    X(x,y) \equiv \Omega(x,y) \ell \sqrt{\left| \frac{P(0)}{P(y)}\right|}.
\end{eqnarray} 
The condition $P(0)>0$ immediately holds for Class~II, but implies that $A\ell<1$ for Classes I and III. Imposing the condition $P(0)>0$, $P(0)=0$, and $P(0)<0$ to Class~I  is respectively equivalent to having the Class $\mathrm{I}_\mathrm{slow}$, Class $\mathrm{I}_\mathrm{saturated}$, and Class $\mathrm{I}_\mathrm{rapid}$ in \cite{Arenas_Henriquez_2022}. 

We note for Class I$_\mathrm{slow}$ that for any given $x$, the maximum value of $|X|$ is
\begin{equation}\label{XmaxI}
    |X|_{max} = \sqrt{1-m^2{\cal A}^2\ell^2(1-x^2)} \leq 1,
\end{equation}
which occurs at 
\begin{equation}\label{rmaxCI}
    y_{max} = \frac{1-1/A^2\ell^2}{x},
\end{equation}
where the inequality, Eq.~\eqref{XmaxI}, is saturated at $|x|=1$. 

Of course, the original C-metric, Eq.~\eqref{cmet}, is not a solution to the field equations, Eq.~\eqref{Eabsimp}, of 3DEGB gravity. However, the result Eq.~\eqref{phiC} provides guidance for obtaining exact solutions. Inserting Eq.~\eqref{phiC} into Eq.~\eqref{Eabsimp}, and expanding in powers of $\alpha$ by writing $g_{ab}=g_{ac}^C(\delta^c{}_b+\alpha h^c{}_b+\mathcal{O}(\alpha^2))$), we find a class of solutions for $h^c{}_d$ whose components are strictly functions of $X$.

This observation naturally motivates the following field and metric ansatz. For the scalar field, we assume
 \begin{eqnarray}
    \phi(x,y)\equiv\phi\left(X(x,y)\right) = \phi(X),
\end{eqnarray}
where $X(x,y)$ is the  dimensionless combination of the prolate coordinates given in Eq.~\eqref{Xdef}. For the metric, we assume the form
\begin{eqnarray}
   ds^2 = \frac{1}{\Omega(x,y)^2} \left[ -F_0(X) P(y) \, d\tau^2 
   + \frac{F_1(X)}{Q(x)} \, dx^2 
   + \frac{F_2(X)}{P(y)} \, dy^2 \right],
  \label{metans} 
\end{eqnarray}
where $F_{0,1,2}(X) > 0$ are functions of $X$; we refer to them as metric factors. These metric factors must be positive to preserve the Lorentzian signature of the metric.

\subsection{Solution strategy and classification of Subclasses\label{subsecclass}}

We now outline our approach for finding new solutions. We first note that $\mathcal{E}_{xx} =\mathcal{E}_{yy} = 0$ in Eq.~\eqref{Eabsimp} requires $F_1(X) = F_2(X) \equiv F(X)$. This identity can be interpreted as the spacetime structure on a fixed time slice being conformally equivalent to that of a C-metric time slice.
Inserting Eq.~\eqref{metans} into Eq.~\eqref{phieq} yields two kinds of solutions, each unique up to a constant. One is
\begin{eqnarray}
\phi(X)=\frac{1}{2}\log\left[\left(\frac{1}{X^2}\pm 1
\right)F(X)\right],\label{phialt}
\end{eqnarray}
but this cannot provide a solution for $\mathcal{E}_{xy} = 0$.

We therefore focus on the other solution
\begin{eqnarray}
    \phi(X)=\frac{1}{2}\log\left[\frac{F_0(X)}{X^2}\right]\label{phifirst}
\end{eqnarray}
which will yield new exact solutions to Eq.~\eqref{Eabsimp}.
The condition $\mathcal{E}_{xy}=0$ then implies
\begin{eqnarray}
    F(X)=\mathfrak{c} \left( \frac{F_0(X)-\tfrac{1}{2}X \partial_X F_0(X)}{\sqrt{F_0(X)}} \right),\label{FKF0}
\end{eqnarray}
where $\mathfrak{c}$ is a positive constant that can be set to unity without loss of generality since it only rescales the time component of the metric.  We can further simplify Eq.~\eqref{FKF0} by writing  
\begin{eqnarray}
    F_0(X)\equiv \frac{X^2}{H(X)^2} \label{defHX},
\end{eqnarray}
yielding
\begin{eqnarray}
F(X) =\frac{X^2}{H(X)^2} \partial_X H(X).
\end{eqnarray}
This, in turn, simplifies the metric ansatz Eq.~\eqref{metans} to the form
\begin{eqnarray}
   ds^2 = \frac{X^2}{\Omega(x,y)^2 H(X)^2} \left[ - P(y) \, d\tau^2 
   + \frac{\partial_X H(X)}{Q(x)} \, dx^2 
   + \frac{\partial_X H(X)}{P(y)} \, dy^2 \right]
  \label{metans2} 
\end{eqnarray}
and the scalar field solution is
\begin{eqnarray}
    \phi(X)=-\log\left|H(X)\right|,
    \label{phi2nd}
\end{eqnarray} 
where $X$ is given in Eq.~\eqref{Xdef}; if $H(X)=X$, we recover the usual C-metric.
We require that $\partial_XH(X)>0$ to preserve the signature. We note that 
\begin{eqnarray}
    g_{\tau\tau}=-\frac{P(y)}{\Omega(x,y)^2}F_0(X)=-\frac{|P(0)|\ell^2}{H(X)^2}.
\end{eqnarray}


Using Eq.~\eqref{FKF0} and our metric ansatz Eq.~\eqref{metans2}, the equations $\mathcal{E}_{yy}=\mathcal{E}_{xx}=0$  can be shown to be equivalent to $\mathcal{E}_{\tau\tau}=0$. The mixed components $\mathcal{E}_{\tau x}$ and $\mathcal{E}_{\tau y}$  vanish identically. 
 Therefore, solving $\mathcal{E}_{xx}=0$ is sufficient to determine the full set of solutions. Finally, we then find that $\mathcal{E}_{xx}=0$ becomes 
\begin{eqnarray}
 H^\prime\left(\frac{2 \alpha}{\ell^2}  s X H-1\right)-\frac{\alpha}{\ell^2}  \left(1+sX^2\right) (H^\prime)^2+  \left(\frac{1+sX H}{1+sX^2}\right) + H H'' \left( \frac{\alpha}{\ell^2}  \left(1+sX^2 \right) +\frac{1}{2H^\prime}\right) = 0,  \label{HX}
\end{eqnarray}
where $H^\prime \equiv\partial_X H$ and
\begin{equation}
    s \equiv \text{sgn} \left(\frac{1}{P(y)}-\frac{1}{P(0)}\right).
\end{equation} 
Hence, solving the metric and scalar field equations reduces to solving Eq.~\eqref{HX}, which is an ordinary differential equation for $H(X)$ only.

We now proceed to carry out this task. Introducing  
\begin{eqnarray}\label{Hint1}
H(X) =H(X_0)+\int_{X_0}^X{\frac{G(Y)}{1 +sY^2}dY},
\end{eqnarray}
we find  that  the  function $G(Y)$ obeys the equation
\begin{eqnarray}\label{Elucidation1}
    2\frac{G(X)}{1+s X^2}\left(1-G(X)-\frac{\alpha}{\ell^2}G(X)^2\right)=\left(H(X_0)+\int_{X_0}^X\frac{G(Y)}{1+s Y^2}dY\right)\left(-1-\frac{2\alpha}{\ell^2}G(X)\right)G'(X)
\end{eqnarray}
as a consequence of Eq.~\eqref{HX}. 

Before solving Eq.~\eqref{Elucidation1}, let us record one useful interpretation of $G(X)$. From the scalar profile $\phi(X)=-\log|H(X)|$, one finds
\begin{equation}
\label{gradphisq}
    (\nabla\phi)^2=\frac{1}{\ell^2}G(X).
\end{equation}
Thus, $G(X)$ controls the scalar-gradient norm. In the polar-coordinate chart introduced in Sec.~\ref{sec:polar}, the same function also determines the proper acceleration of the static origin; see Eq.~\eqref{MagAc}. We now solve for $G(X)$ and classify the resulting branches.

Now, let us solve $G(X)$ and classify possible solutions. We obtain six Subclasses, denoted A, B, C, D, E, and F. We shall delineate all the Subclasses here, but for now, analyze only Subclass A. We will leave a detailed analysis of the remaining subclasses for future work.

\paragraph{Subclasses A and E} 

A simple solution to Eq.~\eqref{Elucidation1} can
be obtained by assuming that both sides of this equation separately vanish. This implies  
\begin{eqnarray}\label{HintA}
    G(X)=B_\alpha \equiv \frac{\ell^2}{2\alpha}\left(-1+\sqrt{1+\frac{4\alpha}{\ell^2}}\right)
    =\frac{2}{\sqrt{1+\frac{4\alpha}{\ell^2}}+1},
\end{eqnarray}
or
\begin{eqnarray}
    G(X)=\frac{\ell^2}{2\alpha}\left(-1-\sqrt{1+\frac{4\alpha}{\ell^2}}\right)
    =-\frac{\ell^2}{\alpha B_\alpha},
\end{eqnarray}
where we note that $B_\alpha \leq 1$, with the inequality saturated for $\alpha=0$. We refer to the choice $G(X)=B_\alpha$ as Subclass A, while the choice $G(X)=-\ell^2/(\alpha B_\alpha)$ as Subclass E.

\paragraph{Subclasses B, C, D, and F}

If $G(X)$ is not a constant, then we can rewrite Eq.~\eqref{Elucidation1} as  
\begin{eqnarray}
    \frac{\frac{d}{dX}\left(H(X_0)+\int_{X_0}^X\frac{G(Y)}{1+s Y^2}dY\right)}{H(X_0)+\int_{X_0}^X\frac{G(Y)}{1+s Y^2}dY}=\frac{1}{2}\frac{\frac{d}{dX}\left(1-G(X)-\frac{\alpha}{\ell^2}G(X)^2\right)}{1-G(X)-\frac{\alpha}{\ell^2}G(X)^2},
\end{eqnarray}
or alternatively,
\begin{eqnarray}\label{HintBC}
H(X_0)+\int_{X_0}^X{\frac{G(Y)}{1 +s Y^2}dY}=\frac{1}{C_1}\sqrt{\left|1-G(X)-\frac{\alpha}{\ell^2}  G(X)^2\right|},\quad C_1\neq 0,
\end{eqnarray}
which is an integral equation for $G(X)$.

There are now two distinct subclasses of solutions depending on the magnitude of $G(X)$. 
If $G(X)>B_\alpha$ or $G(X)<-\frac{\ell^2}{\alpha B_\alpha}$, then Eq.~\eqref{HintBC} is equivalent to 
\begin{eqnarray}\label{HintB}
H(X) =H(X_0)+\int_{X_0}^X{\frac{G(Y)}{1 +s Y^2}dY}=\frac{1}{C_1}\sqrt{-1+G(X)+\frac{\alpha}{\ell^2}  G(X)^2},\quad C_1\neq 0.
\end{eqnarray}
We refer to $G(X)>B_\alpha$ and $G(X)<-\frac{\ell^2}{\alpha B_\alpha}$ as Subclass~B and Subclass~F, respectively. 

However, if $0<G(X)<B_\alpha$ or $-\frac{\ell^2}{\alpha B_\alpha}<G(X)<0$, then Eq.~\eqref{HintBC} is equivalent to 
\begin{eqnarray}\label{HintC}
H(X) =H(X_0)+\int_{X_0}^X{\frac{G(Y)}{1 +s Y^2}dY}=\frac{1}{C_1}\sqrt{1-G(X)-\frac{\alpha}{\ell^2}  G(X)^2},\quad C_1\neq 0.
\end{eqnarray}
We refer to $0<G(X)<B_\alpha$ and $-\frac{\ell^2}{\alpha B_\alpha}<G(X)<0$ as Subclass~C and Subclass~D, respectively.

\paragraph{Domains of the Subclasses}

Before specializing to Subclass~A, it is useful to summarize analytically the branches relevant to the reduced field equation. The sectors $s=+1$, $s=-1$ with $X<1$, and $s=-1$ with $X>1$ should be regarded as distinct branches rather than as different parts of a single smooth static spacetime. The sign of $s$ can change only when $P(y)$ changes sign, namely across the locus $P(y)=0$, where $X\to\infty$. In Subclass~A, two $s=-1$ branches are further separated by the curve $X=1$, and since
\begin{equation}
H'(X)=\frac{B_\alpha}{1-X^2},
\end{equation}
a continuation across $X=1$ fails to be $C^1$. Accordingly, only the branches $s=+1$ and $s=-1$ with $X<1$ are relevant for Subclass~A; the sector $s=-1$ with $X>1$ belongs to Subclasses~D--F and will not be analyzed here.

It is useful to state this restriction class by class. For Class~I,
\begin{equation}
    P(y)=P(0)+y^2,
    \qquad
    P(0)=\frac{1}{A^2\ell^2}-1,
\end{equation}
and therefore, in the timelike region $P(y)>0$,
\begin{equation}
    s=\sgn\!\left(\frac{1}{P(y)}-\frac{1}{P(0)}\right)
    =\sgn\!\left(-\frac{y^2}{P(0)P(y)}\right)
    =-\sgn\!\bigl(P(0)\bigr).
\end{equation}
Hence, $s=-1$ when $A\ell<1$, in which case Eq.~\eqref{XmaxI} implies $0<X\le1$, with $X=1$ attained only at
\begin{equation}
(x,y)=\left(1,-\frac{m}{Ar_{\max}}\right).
\end{equation}
By contrast, when $A\ell>1$, one has $s=+1$, and $X$ can span the full interval $(0,\infty)$.

For Classes~II and III, one has
\begin{equation}
    P(y)=P(0)-y^2.
\end{equation}
Whenever $P(y)>0$, it follows immediately that $P(0)>0$, and so
\begin{equation}
    s=\sgn\!\left(\frac{1}{P(y)}-\frac{1}{P(0)}\right)
    =\sgn\!\left(\frac{y^2}{P(0)P(y)}\right).
    =+1.
\end{equation}
Thus, in the timelike regions of Classes~II and III, only the $s=+1$ branch occurs, and $X$ again ranges over $(0,\infty)$.

In Subclass A, the function $G(X)=B_\alpha$ integrates to
\begin{equation}
    H(X)=H(X_0)+B_\alpha\int_{X_0}^X\frac{dY}{1+sY^2}.
\end{equation}
It is convenient to absorb the $X_0$-dependent contribution into an integration constant $H_0$, defined by
\begin{equation}
    H_0 \equiv
    \begin{cases}
        H(X_0)-B_\alpha\tan^{-1}X_0, & s=+1,\\
        H(X_0)-B_\alpha\tanh^{-1}X_0, & s=-1,\ X<1.
    \end{cases}
\end{equation}
Hence,
\begin{eqnarray}\label{HA}
    H(X)=
    \begin{cases}
        H_0+B_{\alpha}\tan^{-1}X, & s=+1,\\
        H_0+B_{\alpha}\tanh^{-1}X, & s=-1, X<1,
    \end{cases}
\end{eqnarray}
where this generalizes the C-metric classes listed in Table~\ref{table::ClassesPolar}. We note that these solutions
exist and are well defined in the $\alpha\to 0$ limit, and hence represent new solutions to the Einstein-AdS equations in $(2+1)$ dimensions.

In prolate coordinates, the metric of Subclass A is 
\begin{equation}
    ds^2= \frac{1}{\Omega(x,y)^2}\frac{X^2}{H(X)^2}\left(-P(y)d\tau^2+\frac{B_\alpha}{1+sX^2}\frac{dy^2}{P(y)}+\frac{B_\alpha}{1+sX^2}\frac{dx^2}{Q(x)}\right).
\end{equation}
As shown in Appendix~A, this spacetime admits an explicit embedding both in the Poincar\'e patch and in global AdS$_3$.

\section{\label{sec:ana}Analysis of Subclass A}

Henceforth, we restrict attention to the case $H_0=0$. This choice ensures that the metric factors depend only on $X^2$, $H(X)^2$, and $H'(X)=B_\alpha/(1+sX^2)$ in a way that preserves the $(x,y)\to(-x,-y)$ symmetry of the underlying C-metric. Indeed, under $(x,y)\to(-x,-y)$, both $X^2$ and $H'(X)$ remain unchanged, whereas $H(X)^2$ is invariant precisely when $H_0=0$. We may therefore continue to impose the convention $x>y$, equivalently $X>0$ as defined in Eq.~\eqref{Xdef}. Moreover, when $H_0=0$, the metric factors remain finite at $x=y$ (that is, at $X=0$), so the conformal boundary coincides with that of the original C-metric. The more general case $H_0\neq0$ will be discussed in a forthcoming paper.

\subsection{Scalar field behavior}

From Eqs.~\eqref{phi2nd} and \eqref{HA}, the scalar field in Subclass~A is
\begin{equation}
\phi(X)=
\begin{cases}
-\log\!\left(B_\alpha\tan^{-1}X\right), & s=+1,\\[1ex]
-\log\!\left(B_\alpha\tanh^{-1}X\right), & s=-1,\ X<1.
\end{cases}
\label{eq:phiAprofiles}
\end{equation}
In the $s=+1$ branch, $\phi(X)\to\infty$ as $X\to 0$ and asymptotically approaches $-\log \frac{\pi B_\alpha}{2}$ as $X\to \infty$. On the other hand, in the $s=-1$, $X<1$ branch, $\phi(X)\to\infty$ as $X\to 0$ and asymptotically approaches $-\infty$ as $X\to 1$.

\subsection{Metric representations}

In prolate coordinates $(\tau,x,y)$, for the $s=+1$ case, the metric is
\begin{align}\
g_{ab} &= \left(\frac{1}{B_\alpha\tan^{-1}\Xprolate}\right)^2& \nonumber\\
&\times \begin{bmatrix}
-\ell^2|P(0)| & 0 & 0\\
0 & \left({\left|\frac{P(y)}{\ell^2P(0)}\right|+\Omega(x,y)^2}\right)^{-1}\frac{B_\alpha}{P(y)} & 0 \\
0 & 0 & \left({\left|\frac{P(y)}{\ell^2P(0)}\right|+\Omega(x,y)^2}\right)^{-1}\frac{B_\alpha}{Q(x)}
\end{bmatrix}.
\label{eq53}
\end{align}
On the other hand, for the $s=-1$ case, assuming $X<1$, the metric is
\begin{align}\label{minusprolatemetricA}
g_{ab} &=\left(\frac{1}{B_\alpha\tanh^{-1}\Xprolate}\right)^2& \nonumber\\
&\times\begin{bmatrix}
-\ell^2|P(0)| & 0 & 0\\
0 & \left({\left|\frac{P(y)}{\ell^2P(0)}\right|-\Omega(x,y)^2}\right)^{-1}\frac{B_\alpha}{P(y)} & 0 \\
0 & 0 & \left({\left|\frac{P(y)}{\ell^2P(0)}\right|-\Omega(x,y)^2}\right)^{-1}\frac{B_\alpha}{Q(x)}
\end{bmatrix}.
\end{align}
It is straightforward to verify that each of these spacetimes has constant negative curvature, with    
\begin{eqnarray}\label{Rscal}
 R_{ab} =-\frac{2 B_{\alpha}}{\ell^2}g_{ab}  \Rightarrow  R=-\frac{6}{\ell^2}B_{\alpha}=-\frac{6}{\ell_\alpha^2},
\end{eqnarray}
where $\ell_\alpha \equiv \ell/\sqrt{B_\alpha}$; in the $\alpha\to0$ limit, $B_\alpha\to 1$, $\ella\to \ell$, as expected. The effect of $B_\alpha$ is simply to rescale the AdS length $\ell$ from its value in Einstein gravity.

We thus observe that the metric, Eq.~\eqref{minusprolatemetricA}, in the limit $\alpha\to 0$ constitutes a new locally AdS solution to the (2+1)-dimensional Einstein equations. We explore further its properties below.

\subsection{\label{sec:spe}Special limits of Subclass A}

\paragraph{$\alpha=0$}

As noted above, if $\alpha=0$, then the metrics, Eqs.~\eqref{eq53} and \eqref{minusprolatemetricA}, are solutions to Einstein gravity in $(2+1)$ dimensions, where the solution is
\begin{equation}  
 H(X)=
 \begin{cases}\label{a0limit}
  \tan^{-1}(X), &  s=+1,\\
     \tanh^{-1}(X),& s=-1, X<1.
 \end{cases}
\end{equation}
 Both the $H(X)$ in Eq.~\eqref{a0limit} give a locally AdS with cosmological length scale $\ell$. In the small $A$ limit,  $X\to 0$, and therefore, $H(X)\to X$, which recovers the C-metric Eq.~\eqref{cmet}.

\paragraph{Class~I$_\mathrm{saturated}$}

When $A\ell\to 1$ in Class I, we have the limit $P(y)\to y^2$, $P(0)\to0$, and $P(y)\to y^2$. The variable $X=A\ell (x-y)\sqrt{\left|P(0)/P(y)\right|}$   reduces to zero and the solution Eq.~\eqref{HA} is not well-defined.

However, we can obtain solutions in this case by assuming that the metric factors $F_0$, $F_1$, $F_2$, and the scalar field, are instead dependent on  $\bX(x,y) \equiv \ell \Omega(x,y)\sqrt{\left|1/P(y)\right|}=(x-y)/|y|$. 
Assuming that $\alpha\neq0$, we then obtain relations analogous to the $P(0)\neq0$ case: 
\begin{align}
    \phi(\bX)&=-\log |H(\bX)|+\text{const.},\\
    F_0(\bX)&=\frac{
    \bX^2}{H(\bX)^2},\\
    F_1(\bX)&=F_2(\bX)=F(\bX)=\frac{
    \bX^2}{H(\bX)^2}H'(\bX),
\end{align}
where $H(\bX)$ is obtained by solving 
\begin{eqnarray}
    0=\partial_{\bX} H +\frac{\alpha}{\ell^2}  (\partial_{\bX} H)^2-1- H \partial^2_{\bX} H \left( \frac{\alpha}{\ell^2}  +\frac{1}{2\partial_\bX H}\right).\label{HXSat}
\end{eqnarray}
Eq.~\eqref{HXSat} possesses the same form as the $X \to 0$ specialization of Eq.~\eqref{HX}, obtained by dropping all explicit factors of $X$ and $X^{2}$, while retaining the $X$-dependence of $H$ and $\partial_{X} H$, with $\bX$ in place of $X$.

By expressing $H(\bar{X})=H(\bX_0)+\int_{\bX_0}^\bX G(\bX) d\bX$, we can follow the same procedure as in Sec.~\ref{subsecclass}. We obtain different classes depending on  $G(X)$.  If $G(X)=B_\alpha$, then $H(X)=H_0+B_\alpha \bX=\left(H_0+B_\alpha\left(x-y)/|y|\right)\right)$. To preserve the $(x,y)\to(-x,-y)$ symmetry, and hence retain the convenient convention $x>y$ and $\bX>0$, we again set $H_0=0$, leaving the more general case $H_0\neq0$ for future work. The solution for the metric in prolate coordinates is
\begin{align}
    g_{ab}&=\frac{1}{\Omega(x,y)^2}
    \frac{\bX (x,y)^2}{H(\bX (x,y))^2}
\begin{bmatrix}
-P(y) & 0 & 0\\
0 & 1/P(y) & 0\\
0 & 0 & 1/Q(x)
\end{bmatrix}
\begin{bmatrix}
1 & 0 & 0\\
0 & G(\bX (x,y)) & 0\\
0 & 0 & G(\bX (x,y))\\
\end{bmatrix}\\
&=\frac{\ell^2}{B_\alpha^2\left(x-y\right)^2}
\begin{bmatrix}
-y^2 & 0 & 0\\
0 & B_\alpha/y^2 & 0\\
0 & 0 & B_\alpha/(1-x^2)\label{SatprolateA}
\end{bmatrix}.
\end{align}
This is simply the saturated C-metric with all components rescaled by an overall constant factor. We conclude that, in the limit $\alpha\to0$, the standard saturated C-metric is recovered.

\section{\label{sec:polar}Polar coordinate representation and domain conventions}

The solutions are derived most economically in prolate coordinates $(\tau,x,y)$, but the defect interpretation is clearer in polar-type coordinates $(\tau,r,\theta)$. We therefore collect here the coordinate transformation, the polar form of the metric, and the conventions used in the domain plots. This keeps the class-by-class discussion below from repeatedly redefining the same notation.

\subsection{\label{subsec:polar-map}From prolate to polar coordinates}

The relationship between the prolate coordinates $(\tau,x,y)$ and the polar-type coordinates $(\tau,r,\theta)$ is
\begin{subequations}
\label{eqn::polars}
\begin{align}
    y&=-\frac{m}{Ar}=-\frac{1}{\mathcal{A}r},\\
    x&=
    \begin{cases}
        \cos(m\theta), & \mathrm{I}_{\mathrm{wall}},\\
        -\cos(m\theta), & \mathrm{I}_{\mathrm{strut}},\\
        \cosh(m\theta), & \mathrm{II}_{\mathrm{right}},\\
        -\cosh(m\theta), & \mathrm{II}_{\mathrm{left}},\\
        \sinh(m\theta), & \mathrm{III},
    \end{cases}
\end{align}
\end{subequations}
where $\mathcal{A} \equiv A/m$.

For Classes~I and II, $\theta$ is treated as an angular coordinate with range $\theta\in(-\pi,\pi)$, and the identified surfaces $\theta=\pm\pi$ support the wall or strut interpretation. For Class~III, however, the coordinate $x=\sinh(m\theta)$ should be regarded as a hyperbolic polar-type coordinate rather than an angular identification. We include the Class~III row below solely to complete the coordinate dictionary. We do not impose a $\theta=\pm\pi$ identification for Class~III, and we do not include Class~III in the single-defect wall/strut analysis.

In these coordinates, the C-metric takes the form
\begin{equation}
\label{eq:cmetric-polar}
    ds_C^2=
    \frac{1}{\tilde{\Omega}(r,\theta)^2}
    \left[
    -\frac{f(r)}{A^2m^2}\,d\tau^2
    +\frac{dr^2}{f(r)}
    +r^2d\theta^2
    \right].
\end{equation}
The functions $f(r)$ and $\tilde{\Omega}(r,\theta)$ are listed in Table~\ref{table::ClassesPolar}. The allowed range of $r$ is obtained by rewriting the prolate-coordinate condition $x>y$ in terms of $r$ and $\theta$.

\begin{table}[H]
\centering
\caption{Polar coordinate data for the Class~I and Class~II C-metric branches.}
\begin{tabular}{|c|c|c|c|c|} 
\hline
Class & $f(r)$ & $\tilde{\Omega}(r,\theta)$ & Range of $m$ & Range of $r$ \\ 
\hline\hline
I$_\mathrm{wall}$ 
& $\frac{r^2}{\ell^2}+m^2(1-\mathcal{A}^2r^2)$ 
& $1+\mathcal{A}r\cos(m\theta)$ 
& $0\leq m<1$ 
& $\frac{1}{\mathcal{A}r}>-\cos(m\theta)$
\\[0.5ex]
I$_\mathrm{strut}$ 
& $\frac{r^2}{\ell^2}+m^2(1-\mathcal{A}^2r^2)$ 
& $1-\mathcal{A}r\cos(m\theta)$ 
& $0\leq m<1$ 
& $\frac{1}{\mathcal{A}r}>\cos(m\theta)$
\\[0.5ex]
II$_\mathrm{right}$ 
& $\frac{r^2}{\ell^2}-m^2(1-\mathcal{A}^2r^2)$ 
& $1+\mathcal{A}r\cosh(m\theta)$ 
& $m>0$ 
& $\frac{1}{\mathcal{A}r}>-\cosh(m\theta)$
\\[0.5ex]
II$_\mathrm{left}$ 
& $\frac{r^2}{\ell^2}-m^2(1-\mathcal{A}^2r^2)$ 
& $1-\mathcal{A}r\cosh(m\theta)$ 
& $m>0$ 
& $\frac{1}{\mathcal{A}r}>\cosh(m\theta)$
\\[0.5ex]
III 
& $\frac{r^2}{\ell^2}-m^2(1+\mathcal{A}^2r^2)$ 
& $1+\mathcal{A}r\sinh(m\theta)$ 
& $m\in \mathbb{R}$ 
& $\frac{1}{\mathcal{A}r}>-\sinh(m\theta)$
\\[0.5ex]
\hline
\end{tabular}
\label{table::ClassesPolar}
\end{table}

The class-by-class figures in the following sections use a common color convention.
Purple denotes the branch $s=+1$, for which the Killing coordinate $\tau$ is timelike. Light green denotes $s=-1$ with $X<1$ and spacelike $\tau$. Red denotes a region whose metric would be Euclidean if analytically extended from the $s=-1$ region, with   $X<1$. White denotes regions beyond the conformal boundary.
Class~I$_\mathrm{saturated}$ is represented entirely by orange.

\subsection{\label{subsec:origin-acceleration}Static origin and the meaning of $G(X)$}

The polar chart also clarifies the geometric meaning of $G(X)$. A static observer at the origin has a normalized velocity
\begin{equation}
    \mathbf{u}
    =
    Am\frac{\tilde{\Omega}(r,\theta)}
    {\sqrt{F_0(X)f(r)}}\,
    \partial_\tau
    \Big|_{r\to0},
\end{equation}
where $F_0(X)=X^2/H(X)^2$. The associated acceleration is
\begin{equation}
    \mathbf{a}
    =
    \nabla_{\mathbf{u}}\mathbf{u}\big|_{r\to0}.
\end{equation}
Its magnitude is
\begin{equation}
\label{MagAc}
    |\mathbf{a}|
    =
    \frac{1}{\ell}\sqrt{|G(X)|}\Big|_{r\to0}
    =
    |\nabla\phi(X)|\Big|_{r\to0}.
\end{equation}
Moreover,
\begin{equation}
    \lim_{r\to0}X=A\ell\sqrt{|P(0)|}.
\end{equation}
Therefore, $G(X)$ determines the acceleration of the static origin. For the original C-metric, $H(X)=X$, so $G(X)=1+sX^2$. In Class~I this gives
\begin{equation}
    |\mathbf{a}|_{\mathrm{C\text{-}metric}}
    =
    \frac{1}{\ell}
    \sqrt{\left|1+sA^2\ell^2|P(0)|\right|}
    =
    A,
\end{equation}
as expected.

\subsection{\label{subsec:polar-metric}Polar form of the Subclass A metric}

For the general Subclass A solution, the prolate coordinate expression can be written in polar coordinates as
\begin{equation}
\label{eq:subclassA-polar-compact}
    ds_A^2=
    \frac{X(r,\theta)^2}{\tilde{\Omega}(r,\theta)^2H(X)^2}
    \left[
    -\frac{f(r)}{A^2m^2}\,d\tau^2
    +H'(X)
    \left(
    \frac{dr^2}{f(r)}
    +r^2d\theta^2
    \right)
    \right],
\end{equation}
where
\begin{equation}
\label{eq:X-polar}
    X(r,\theta)
    =
    \left|
    A\ell m\,\tilde{\Omega}(r,\theta)
    \sqrt{\frac{P(0)}{f(r)}}
    \right|.
\end{equation}
In Subclass A,
\begin{equation}
\label{eq:Hprime-subclassA}
    H'(X)=\frac{B_\alpha}{1+sX^2},
\end{equation}
with
\begin{equation}
\label{eq:H-subclassA-polar}
    H(X)=
    \begin{cases}
        B_\alpha\tan^{-1}X, & s=+1,\\[0.5ex]
        B_\alpha\tanh^{-1}X, & s=-1,\ X<1,
    \end{cases}
\end{equation}
where we have imposed $H_0=0$. The compact expression, Eq.~\eqref{eq:subclassA-polar-compact}, is the form used conceptually below. For the domain plots, however, it is useful to explicitly display the two relevant component forms.

For the $s=+1$ branch of Subclass A, substituting $H(X)=B_\alpha\tan^{-1}X$ into Eq.~\eqref{eq:subclassA-polar-compact} gives
\begin{align}
\label{C23A}
    g_{ab}^{(+)}
    &=
    \left(
    \frac{1}{B_\alpha\tan^{-1}\Xpolar}
    \right)^2\\&
    \begin{bmatrix}
        -\ell^2|P(0)|\sgn(f(r)) & 0 & 0\\[0.5ex]
        0 &
        \left(
        \left|
        \frac{f(r)}{A^2\ell^2m^2P(0)}
        \right|
        +\tilde{\Omega}(r,\theta)^2
        \right)^{-1}
        \frac{B_\alpha}{f(r)}
        & 0\\[1ex]
        0 & 0 &
        \left(
        \left|
        \frac{f(r)}{A^2\ell^2m^2P(0)}
        \right|
        +\tilde{\Omega}(r,\theta)^2
        \right)^{-1}
        B_\alpha r^2
    \end{bmatrix}.\notag
\end{align}
This is the branch that appears in the timelike regions of Classes II and III, and also in the rapid Class I sector.

For the $s=-1$ branch of Subclass A, restricted to $X<1$, one instead has $H(X)=B_\alpha\tanh^{-1}X$. Before simplifying, the metric can be written as
\begin{align}
\label{eq:sminus-polar-unsimplified}
    g_{ab}^{(-)}
    &=
    \frac{1}{\tilde{\Omega}(r,\theta)^2}
    \left(
    \frac{X}{B_\alpha\tanh^{-1}X}
    \right)^2
    \begin{bmatrix}
        -\frac{f(r)}{A^2m^2} & 0 & 0\\[0.5ex]
        0 & \frac{B_\alpha}{1-X^2}\frac{1}{f(r)} & 0\\[0.5ex]
        0 & 0 & \frac{B_\alpha}{1-X^2}r^2
    \end{bmatrix}.
\end{align}
Using Eq.~\eqref{eq:X-polar}, this becomes
\begin{align}
\label{C1A}
    g_{ab}^{(-)}
    &=
    \left(
    \frac{1}{B_\alpha\tanh^{-1}\Xpolar}
    \right)^2\\
    &
    \begin{bmatrix}
        -\ell^2|P(0)|\sgn(f(r)) & 0 & 0\\[0.5ex]
        0 &
        \left(
        \left|
        \frac{f(r)}{A^2\ell^2m^2P(0)}
        \right|
        -\tilde{\Omega}(r,\theta)^2
        \right)^{-1}
        \frac{B_\alpha}{f(r)}
        & 0\\[1ex]
        0 & 0 &
        \left(
        \left|
        \frac{f(r)}{A^2\ell^2m^2P(0)}
        \right|
        -\tilde{\Omega}(r,\theta)^2
        \right)^{-1}
        B_\alpha r^2
    \end{bmatrix}.\notag
\end{align}
The restriction $X<1$ is equivalent to
\begin{equation}
    \tilde{\Omega}(r,\theta)^2
    <
    \left|
    \frac{f(r)}{A^2\ell^2m^2P(0)}
    \right|.
\end{equation}
For the metrics in Eq.~\eqref{C23A}, relevant to Classes II and III in the appropriate regions, the conformal factor is everywhere nonzero and finite. 

For both metrics Eqs.~\eqref{C23A} and \eqref {C1A}, the function $f(r)$ can vanish 
in Class~I$_\mathrm{rapid}$, Class~II, and Class~III$_\mathrm{slow}$.
This does not correspond to a horizon; at such a locus, $g_{rr}$ diverges while the other metric components remain finite. Rather, this indicates a soliton-like endpoint of the spatial slices at constant $\tau$, rather than a curvature singularity. To see this, define a proper radial coordinate near $f(r)=0$ by
\begin{equation}
    R=\int\frac{dr}{\sqrt{f(r)}}.
\end{equation}
In terms of $R$, the apparent divergence of $g_{rr}$ is the usual coordinate divergence associated with the endpoint of the radial coordinate.

We also note that the metric is static only in regions where the Killing coordinate $\tau$ is timelike. This occurs in both metrics Eqs.~\eqref{C23A} and \eqref{C1A} if $f(r)>0$; otherwise $\tau$ becomes a spacelike coordinate, and the coordinate $r\to T$ becomes timelike. In this case, we obtain new metrics that are non-static patches of AdS, whose spatial sections are $\mathbb{R}\times S^1$. From Table II, 
the restriction $f(T) < 0$ suggests these cases correspond to spacetimes with a big bang and big crunch, though their physical interpretation remains unclear at present.

\section{\label{sec:spa}Spacetime analysis of Subclass A}

We now apply the domain conventions of Sec.~\ref{sec:polar} to the five cases that appear in the Subclass~A solution. Class~I$_\mathrm{slow}$ is the point-particle branch with $A\ell<1$; it remains entirely on the $s=-1$, $X<1$ branch. Class~I$_\mathrm{saturated}$ is the limiting point-particle branch with $A\ell=1$, for which $P(0)=0$ and the special variable $\bar X$ must be used. Class~I$_\mathrm{rapid}$ has $A\ell>1$ and contains disconnected Subclass A branches; in the $r<0$ chart, it is further divided into the \textit{Slightly Rapid} regime, $1<A\ell<\sqrt{2}$, and the \textit{Very Rapid} regime, $A\ell>\sqrt{2}$. Class~II is the accelerating black hole branch, with right and left components corresponding to $x>1$ and $x<-1$, respectively. Class~III is a nonstandard branch; it is included for completeness, but it is not used for a single wall or strut construction.

The colors of the plots label the branches of the solutions to Eq.~\eqref{HX}, rather than different matter contents or spacetime topologies, and indicate the causal character of the Killing coordinate. Purple denotes the branch $s=+1$ with timelike $\tau$. Yellow denotes the branch $s=-1$, $X<1$, with timelike $\tau$. Light green denotes the branch $s=-1$, $X<1$, with spacelike $\tau$.  
Orange denotes the saturated Class~I branch. Red denotes the region whose metric would be Euclidean if analytically extended from the $s=-1$, with the $X<1$ case. White denotes beyond the conformal boundary. 

In the plots below, we impose the timelike condition on $\tau$ separately from the coordinate-domain condition $x>y$. Thus, a colored region may be geometrically present but irrelevant for the static wall or strut interpretation. In the original C-metric, a value $y=y_h>0$ satisfying $P(y_h)=0$ corresponds to an event horizon. In the Subclass A geometries studied here, the same value of $y_h$ is only a useful marker inherited from the C-metric. It should not automatically be interpreted as an event horizon.

The polar-domain figures plot
\begin{equation}
    R_x=|r|\cos\theta,
    \qquad
    R_y=|r|\sin\theta.
\end{equation}
These are plotting variables, not a new global Cartesian coordinate system. Each polar plot should use either the $r>0$ patch or the $r<0$ patch. The two signed $r$ patches should not be superposed in a single panel.

For Class~I$_\mathrm{slow}$, the conformal factor can vanish only when
\begin{equation}
    X
    =
    \left|
    m\tilde{\Omega}(r,\theta)
    \sqrt{
    \frac{1-A^2\ell^2}{f(r)}
    }
    \right|
    =
    1.
\end{equation}
Using Eq.~\eqref{rmaxCI}, this occurs only at
\begin{equation}
    \theta=0,
    \qquad
    r=
    \frac{A^2\ell^2}{1-A^2\ell^2}\,\ell .
\end{equation}
The curvature remains finite and constant at this point, so this locus is not a horizon. It is instead a limitation of the polar coordinate patch.

\subsection{Class I$_\mathrm{slow}$}

\begin{figure}[H]
    \centering
    \includegraphics[scale=0.45]{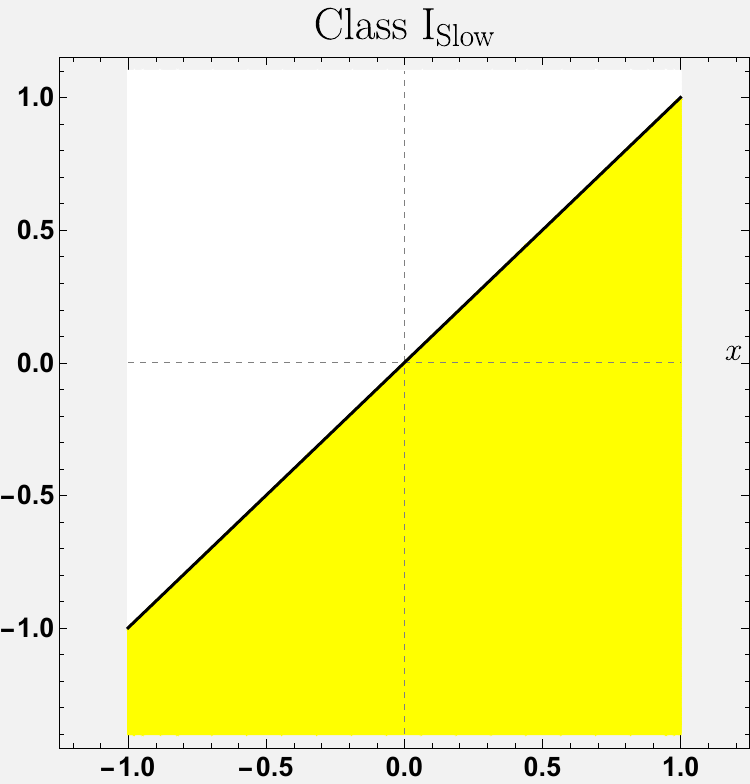}
    \caption{Coordinate domain of Class~I$_\mathrm{slow}$ in the $(x,y)$ plane. The yellow region denotes the branch $s=-1$ with $X<1$ and timelike $\tau$. Since the entire physical domain lies in this single sector, Class~I$_\mathrm{slow}$ is modified uniformly throughout the connected spacetime.}
  \label{fig:class-i-slow-prolate}
\end{figure}

Class~I$_\mathrm{slow}$ is the simplest case because $P(y)>0$ throughout the physical region. As shown in Fig.~\ref{fig:class-i-slow-prolate}, the whole prolate coordinate domain lies in the yellow sector, namely the $s=-1$, $X<1$ branch with timelike $\tau$. Thus, the Subclass~A deformation does not split the spacetime into distinct branches in this case.

\begin{figure}[H]
\centering

\begin{subfigure}{0.47\textwidth}
  \centering
  \includegraphics[scale=0.45]{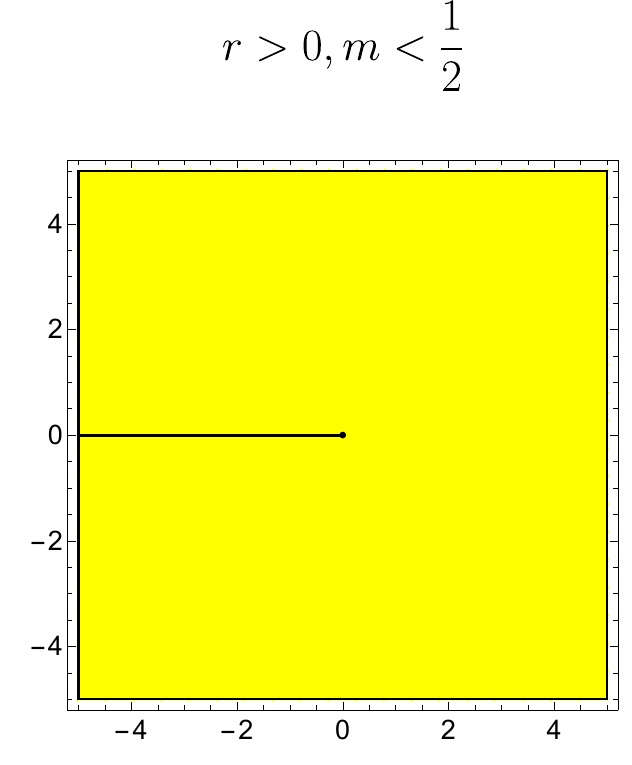}
  \caption{}
\end{subfigure}
\hfill
\begin{subfigure}{0.47\textwidth}
  \centering
  \includegraphics[scale=0.45]{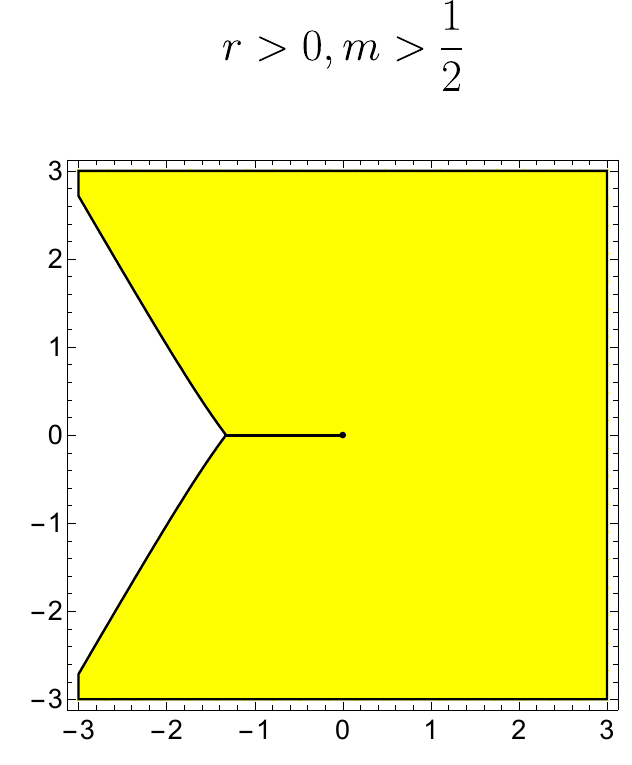}
  \caption{}
\end{subfigure}

\vspace{0.4cm}

\begin{subfigure}{0.47\textwidth}
  \centering
  \includegraphics[scale=0.45]{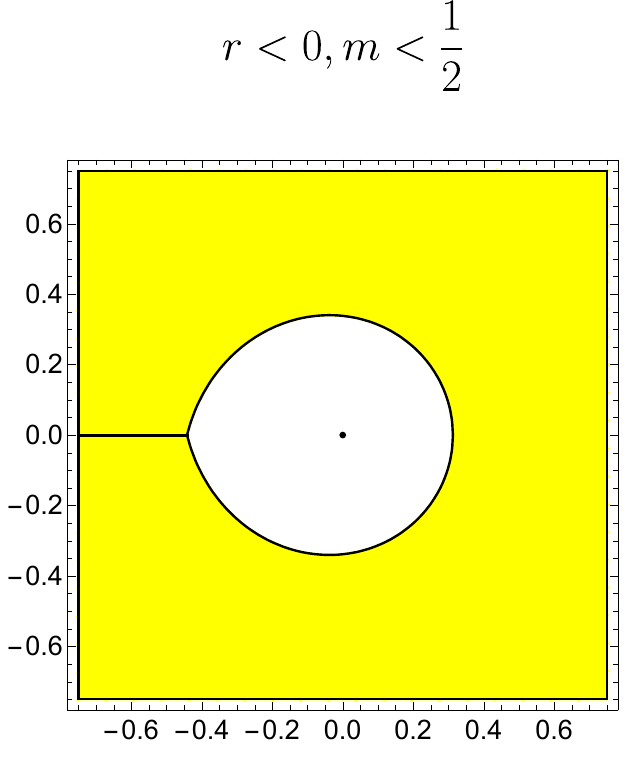}
 \caption{}
\end{subfigure}
\hfill
\begin{subfigure}{0.47\textwidth}
  \centering
  \includegraphics[scale=0.45]{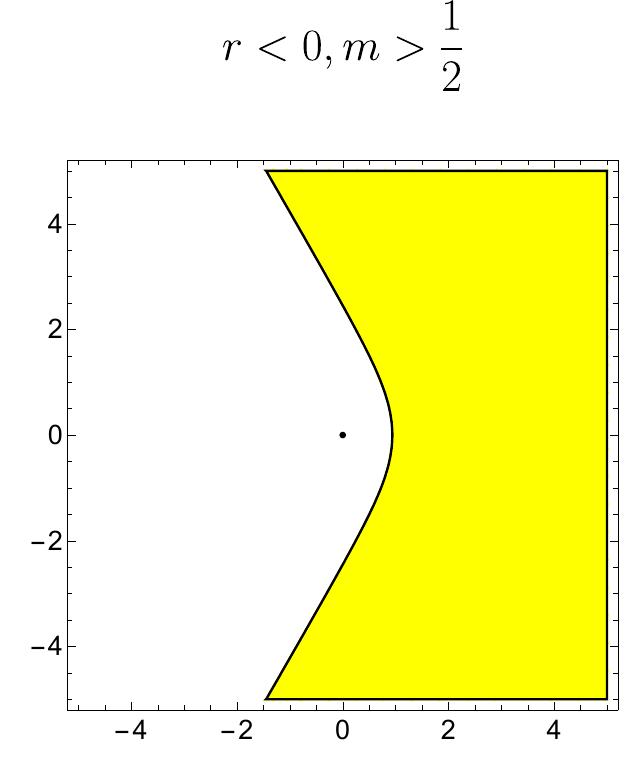}
  \caption{}
\end{subfigure}

\caption{Polar representations of the Class~I$_\mathrm{slow}$ wall branch. Yellow denotes $s=-1$, $X<1$ with timelike $\tau$, and white denotes the excluded region beyond the conformal boundary.
In (c,d), the point $r=0$ lies beyond the conformal boundary, so this chart does not itself contain the origin at which the particle would be located.
}
\label{fig:class-i-slow-wall-polar}
\end{figure}

\paragraph{Wall}
The wall branch is shown in polar form in Fig.~\ref{fig:class-i-slow-wall-polar}. The separate panels display the relevant $r>0$ and $r<0$ charts of the same connected slow Class~I spacetime. Since no $s=+1$ sector appears, every colored region in the figure belongs to the same static branch. The white regions are outside the Subclass A coordinate domain; in the lower panels, this excludes the coordinate point $r=0$, so those charts do not, by themselves, contain the origin at which the particle would be located.

\paragraph{Strut}

\begin{figure}[H]
\centering
\mbox{
(a)
\includegraphics[scale=0.45]{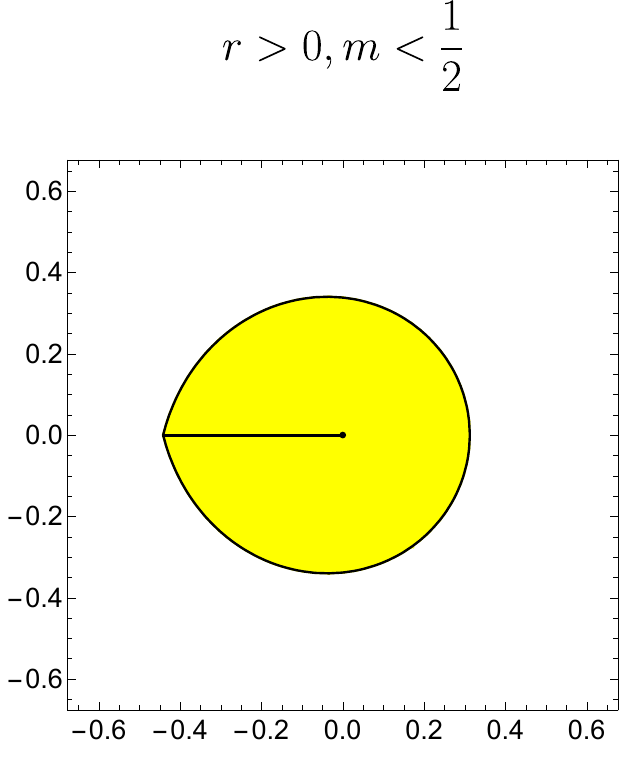}
(b)
\includegraphics[scale=0.45]{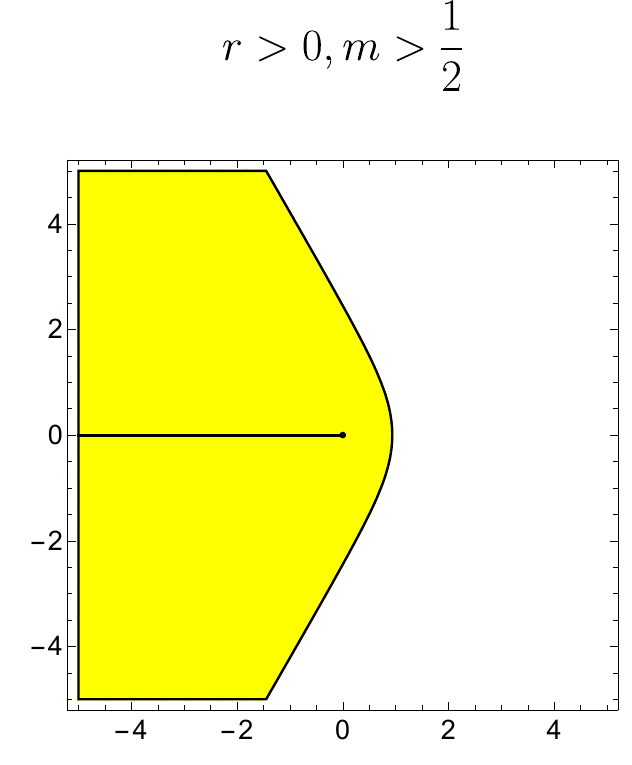}
(c)
\includegraphics[scale=0.45]{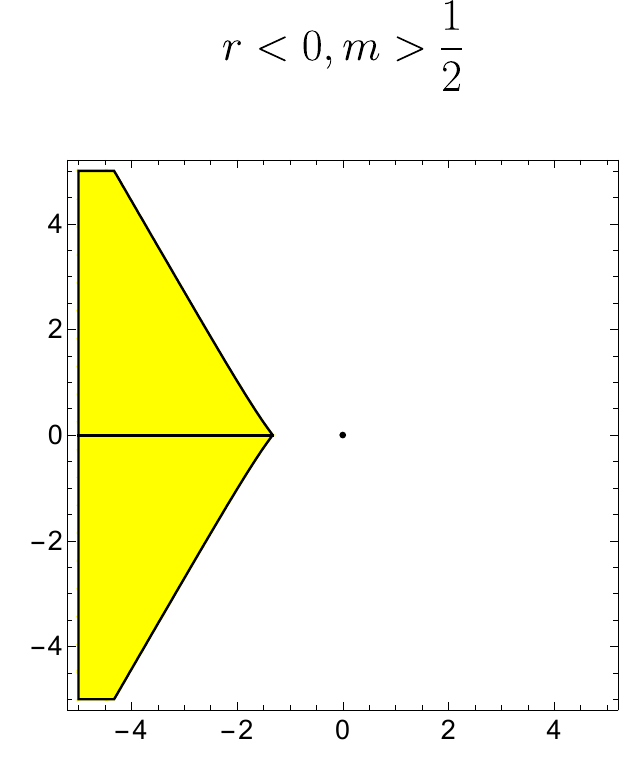}
}
\caption{Polar representations of the Class~I$_\mathrm{slow}$ strut branch. The displayed $r>0$ and $r<0$ charts belong to the same connected spacetime; white regions lie beyond the conformal boundary.}\label{fig:class-i-slow-strut-polar}
\end{figure}

The corresponding strut branch is shown in Fig.~\ref{fig:class-i-slow-strut-polar}. As in the wall case, the $y<0$ ($r>0$) and $y>0$ ($r<0$) regions are connected through the coordinate locus $y=0$, equivalently $r=\pm\infty$. The figure should therefore be read as a set of charts for one connected spacetime, rather than as independent solutions. The only relevant static sector is again the yellow $s=-1$, $X<1$ branch.

\subsection{Class I$_\mathrm{saturated}$}

Turning Eq.~\eqref{SatprolateA} into the polar coordinate, the metric is 
\begin{align}
    g_{ab}&=\frac{1}{\tilde{\Omega}(r,\theta)^2}
    \frac{\bX (r,\theta)^2}{H(\bX (r,\theta))^2}
\begin{bmatrix}
-f(r)/A^2m^2 & 0 & 0\\
0 & 1/f(r) & 0\\
0 & 0 & r^2
\end{bmatrix}
\begin{bmatrix}
1 & 0 & 0\\
0 & G(\bX (r,\theta)) & 0\\
0 & 0 & G(\bX (r,\theta))\\
\end{bmatrix}\\
&=\frac{1}{B_\alpha^2\tilde{\Omega}^2(r,\theta)}
\begin{bmatrix}
-\ell^2 & 0 & 0\\
0 & B_\alpha/m^2 & 0\\
0 & 0 & B_\alpha r^2
\end{bmatrix}.
\end{align}

Each saturated Class~I patch has essentially the same local structure as the corresponding slow Class~I patch. The key difference is that the $r>0$ and $r<0$ branches are now separated, that is,  are not connected within a single static chart. For this reason, the $r<0$ branch cannot be interpreted as being pulled or pushed by the defect in the $r>0$ branch, and there is therefore no saturated analog of the $r<0$, $m<\frac{1}{2}$ case that occurs in Class~I$_\mathrm{slow}$.

\begin{figure}[H]
    \centering
    \includegraphics[scale=0.45]{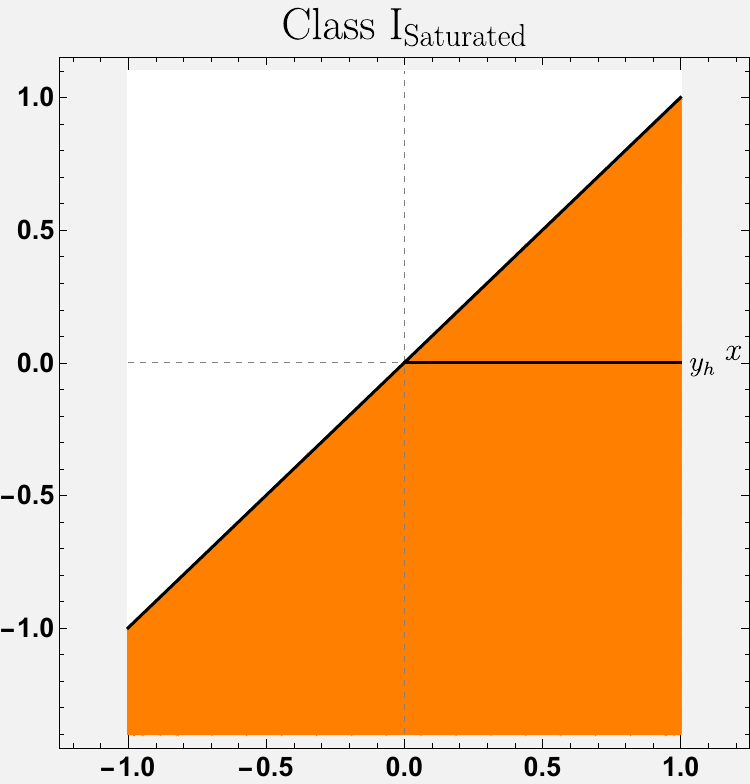}
    \caption{Coordinate domain of Class~I$_\mathrm{saturated}$ in the $(x,y)$ plane. The orange region denotes the saturated branch, which occurs only in this case. Inside the conformal boundary, the Killing coordinate $\tau$ is timelike throughout the physical region and becomes null only at the degenerate horizon $y=0$. The $r>0$ and $r<0$ sectors are therefore treated as distinct static branches. The black horizontal line depicts the event horizon at $y=y_h=0$ (equivalently, $r=\pm \infty$). }
\end{figure}

\begin{figure}[H]
\centering
\mbox{
(a)
\includegraphics[scale=0.45]{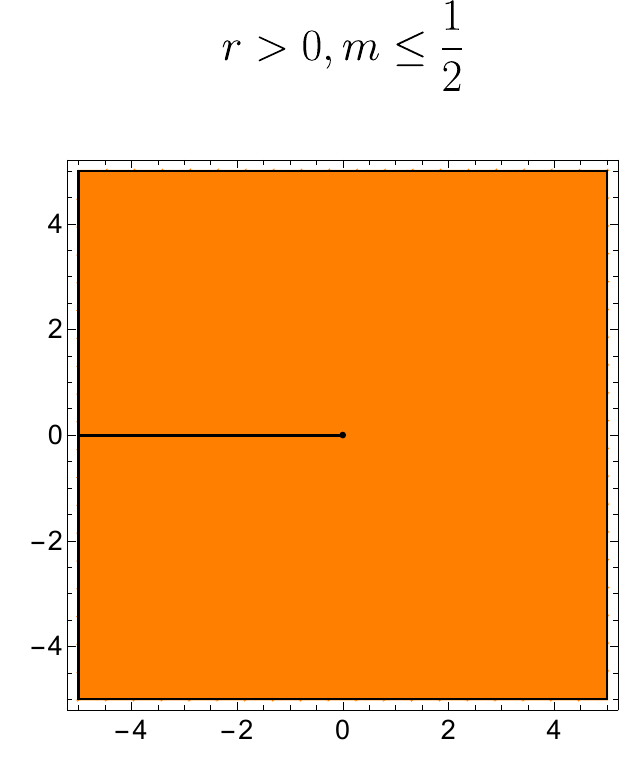}
(b)
\includegraphics[scale=0.45]{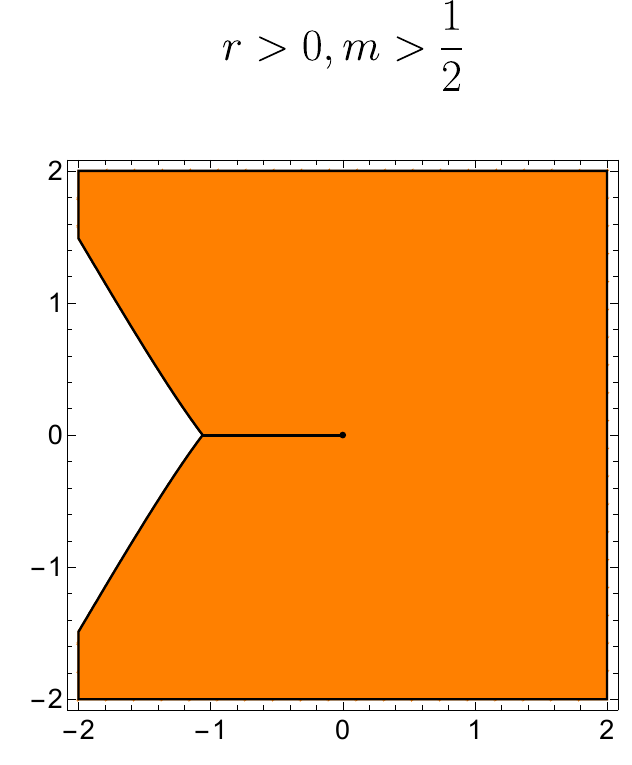}
(c)
\includegraphics[scale=0.45]{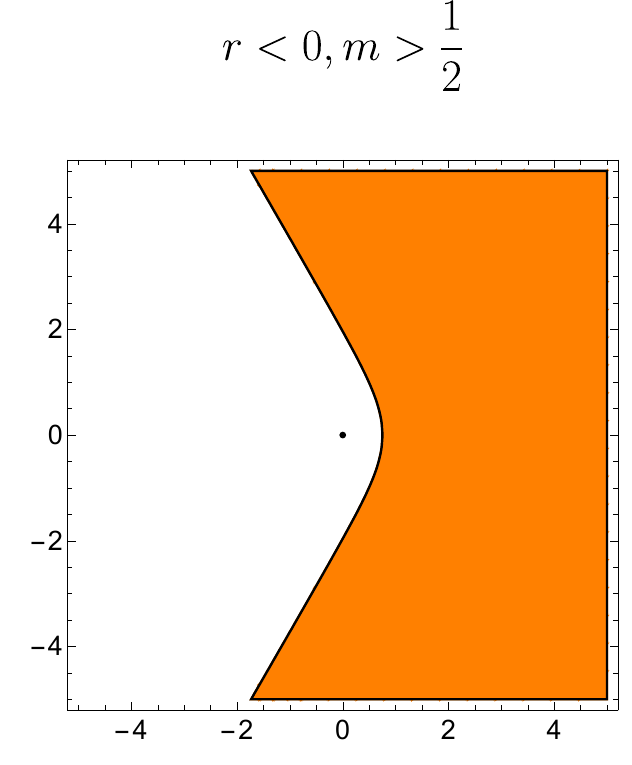}
}
\caption{Polar representations of the Class~I$_\mathrm{saturated}$ wall branch. Orange denotes the saturated timelike sector; white denotes the excluded region beyond the conformal boundary.}
\label{fig:class-i-saturated-wall-polar}
\end{figure}

\paragraph{Wall}
The saturated wall charts are shown in Fig.~\ref{fig:class-i-saturated-wall-polar}. Each patch has the same local structure as the corresponding slow Class~I patch, but the $r>0$ and $r<0$ branches are separated and are not connected within a single static chart. For this reason, the $r<0$ branch cannot be interpreted as being pulled or pushed by the defect in the $r>0$ branch, and there is no saturated analog of the connected $r<0$, $m<\frac{1}{2}$ case that occurs for Class~I$_\mathrm{slow}$.

\paragraph{Strut}

\begin{figure}[H]
\centering
\mbox{
(a)
\includegraphics[scale=0.45]{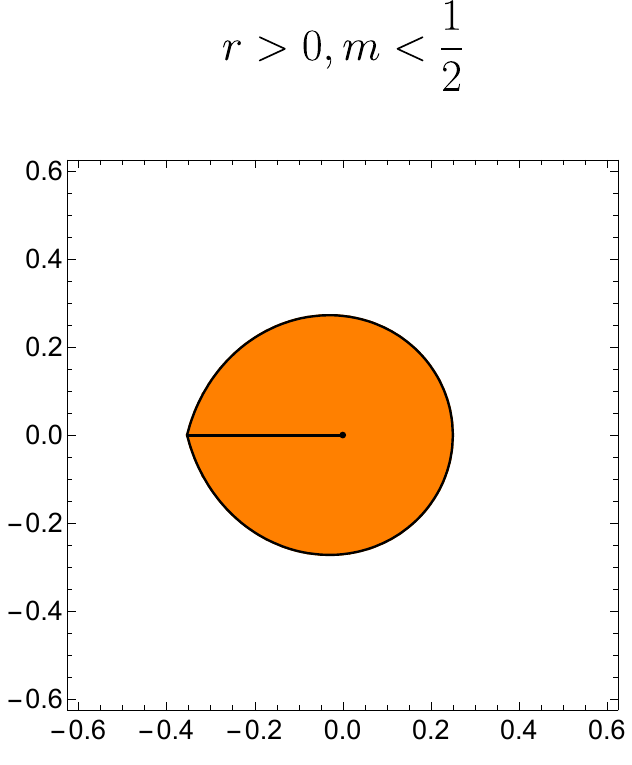}
(b)
\includegraphics[scale=0.45]{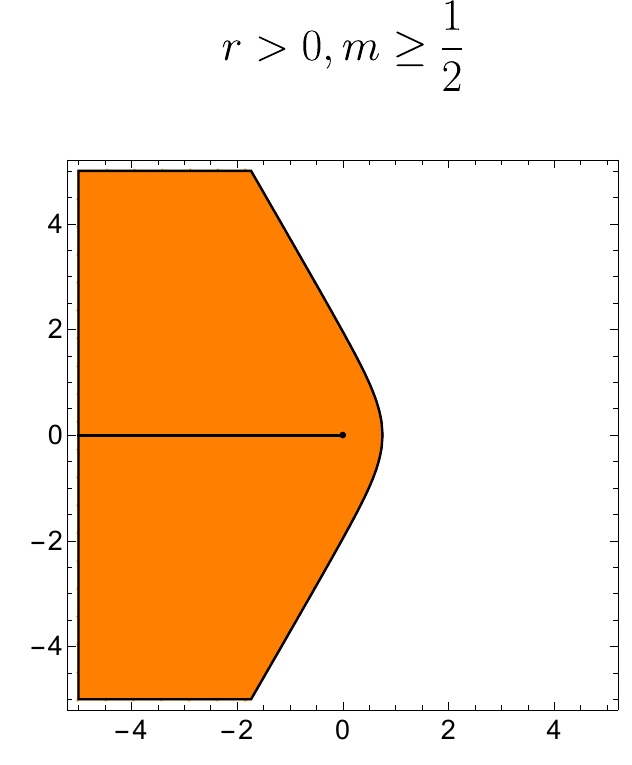}
}
\caption{Polar representations of the Class~I$_\mathrm{saturated}$ strut branch. Orange denotes the saturated timelike sector; white denotes the excluded region beyond the conformal boundary.}\label{fig:class-i-saturated-strut-polar}
\end{figure}

Fig.~\ref{fig:class-i-saturated-strut-polar} shows the saturated strut branch. Only the static branches containing the identified surface at $\theta=\pm\pi$ are relevant for the strut interpretation. As in the wall branch, the saturated limit separates the $r>0$ and $r<0$ sectors, so the figure should not be read as a single connected chart across $r=\pm\infty$.

\subsection{Class I$_\mathrm{rapid}$}

\begin{figure}[H]
    \centering
    \includegraphics[scale=0.45]{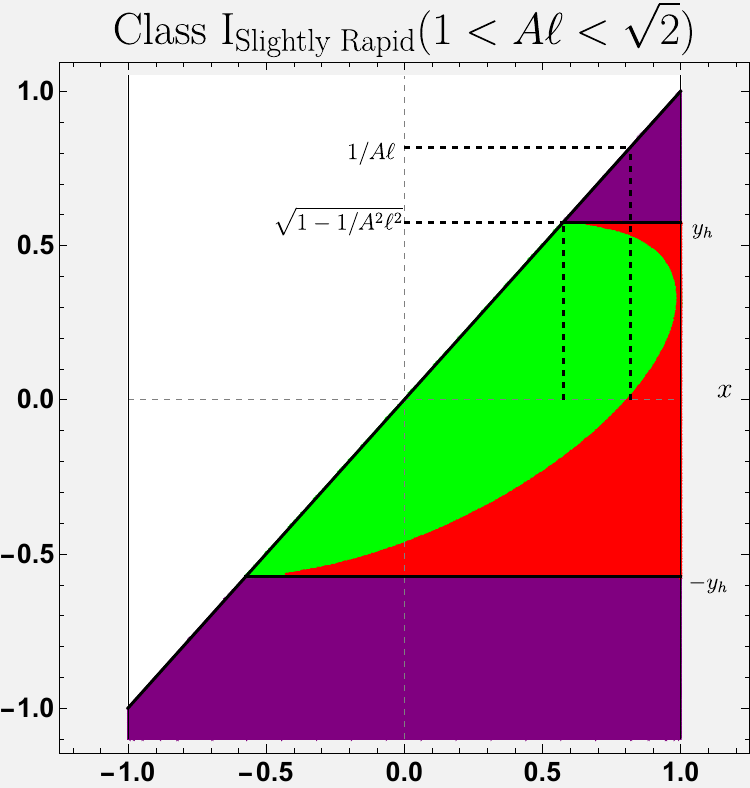}
    \caption{Coordinate domains of Class~I$_\mathrm{rapid}$ in the $(x,y)$ plane for the Slightly Rapid regime, $1<A\ell<\sqrt{2}$. Purple denotes the branch $s=+1$, for which the Killing coordinate $\tau$ is timelike. Light green denotes $s=-1$ with $X<1$ and spacelike $\tau$. Red denotes a region whose metric would be Euclidean if analytically extended from the $s=-1$ region, with   $X<1$. White denotes regions beyond the conformal boundary.}
    \label{fig:class-i-rapid-slight-prolate}
\end{figure}

\begin{figure}[H]
    \centering
    \includegraphics[scale=0.45]{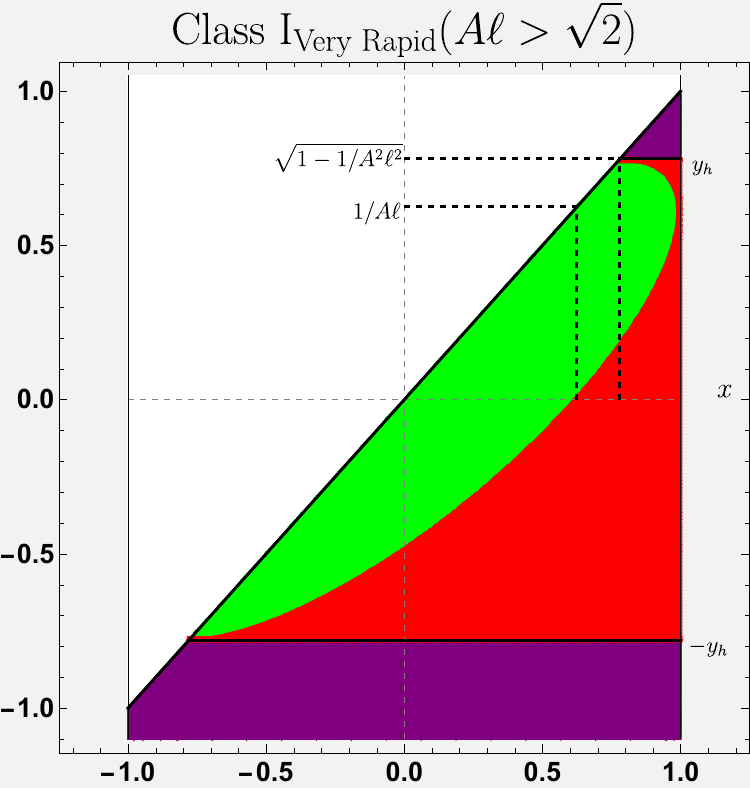}
    \caption{Coordinate domains of Class~I$_\mathrm{rapid}$ in the $(x,y)$ plane for the Very Rapid regime, $A\ell>\sqrt{2}$. The ordering of $x_{X=1}$ and $x_h$ is reversed relative to Fig.~\ref{fig:class-i-rapid-slight-prolate}. Purple denotes the branch $s=+1$, for which the Killing coordinate $\tau$ is timelike. Light green denotes $s=-1$ with $X<1$ and spacelike $\tau$. Red denotes a region whose metric would be Euclidean if analytically extended from the $s=-1$ region, with $X<1$. White denotes regions beyond the conformal boundary.}
    \label{fig:class-i-rapid-very-prolate}
\end{figure}

Class~I$_\mathrm{rapid}$ differs qualitatively from the slow and saturated cases because the $s=+1$ and $s=-1$, $X<1$ sectors can both appear within the coordinate domain. The would-be $s=-1$, $X>1$ sector is outside Subclass~A and is therefore omitted from the Subclass A domain plots. The distinction between the two rapid regimes is visible in Figs.~\ref{fig:class-i-rapid-slight-prolate} and~\ref{fig:class-i-rapid-very-prolate}. The curve $X=1$ intersects the $x$-axis at
\begin{equation}
    x_{X=1}=\frac{1}{A\ell},
\end{equation}
whereas the curve $P(y)=0$, equivalently $f(r)=0$, meets the conformal boundary at
\begin{equation}
    x_h=y_h=\sqrt{1-\frac{1}{A^2\ell^2}}.
\end{equation}
When $1<A\ell<\sqrt{2}$ one has $x_{X=1}>x_h$, which defines the \textit{Slightly Rapid} regime. When $A\ell>\sqrt{2}$ one has $x_{X=1}<x_h$, which defines the \textit{Very Rapid} regime.

\paragraph{Wall}

For the wall interpretation, we retain only connected components that contain the identification surface at $\theta=\pm\pi$ within a static region. Since the $s=+1$ and $s=-1$, $X<1$ sectors correspond to distinct branches of the reduced field equation, they are treated as separate spacetimes rather than as parts of one smooth static solution. The $s=-1$, $X<1$ branch is relevant only where $\tau$ remains timelike. When two charts belong to the same connected branch, it is sufficient that the defect be present in one of them; it need not appear simultaneously in both the $r>0$ and $r<0$ charts.

\paragraph{$r>0$ patch}

\begin{figure}[H]
\centering

\begin{subfigure}{0.47\textwidth}
  \centering
  \includegraphics[scale=0.45]{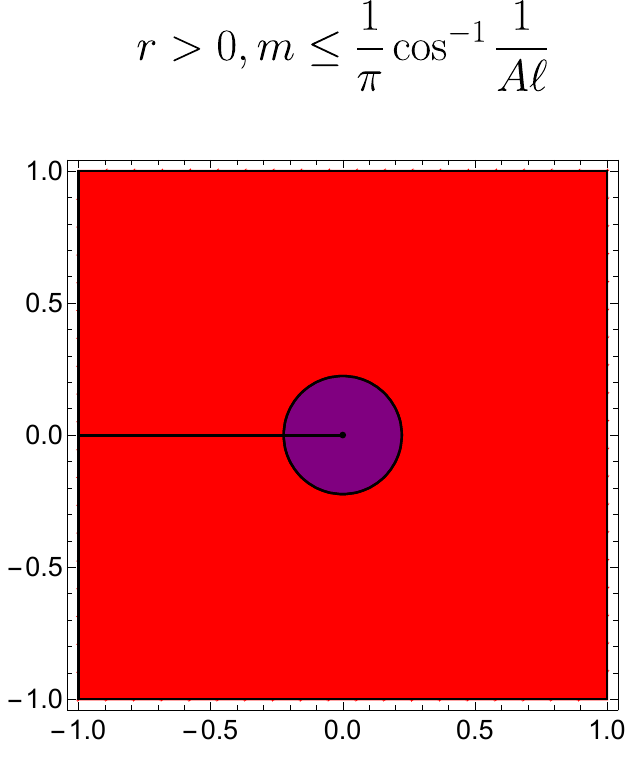}
  \caption{}
\end{subfigure}
\hfill
\begin{subfigure}{0.47\textwidth}
  \centering
  \includegraphics[scale=0.45]{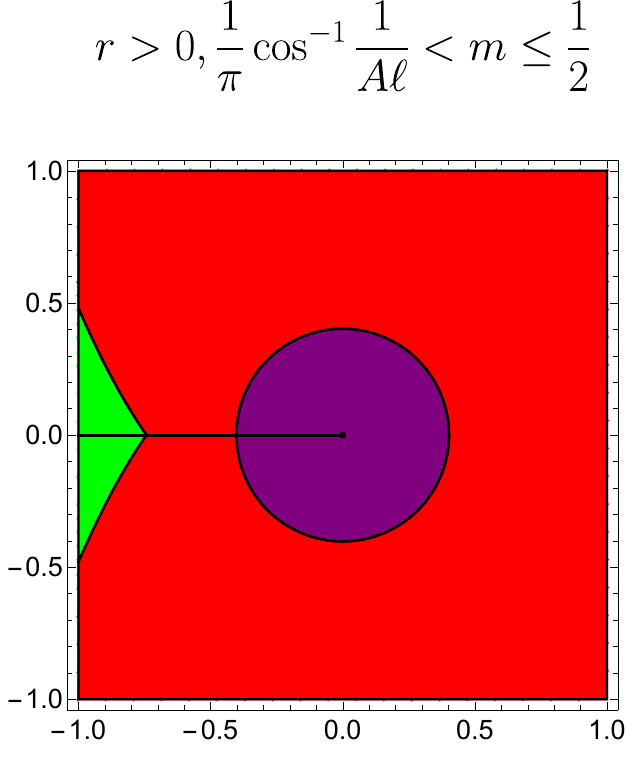}
  \caption{}
\end{subfigure}

\vspace{0.4cm}

\begin{subfigure}{0.47\textwidth}
  \centering
  \includegraphics[scale=0.45]{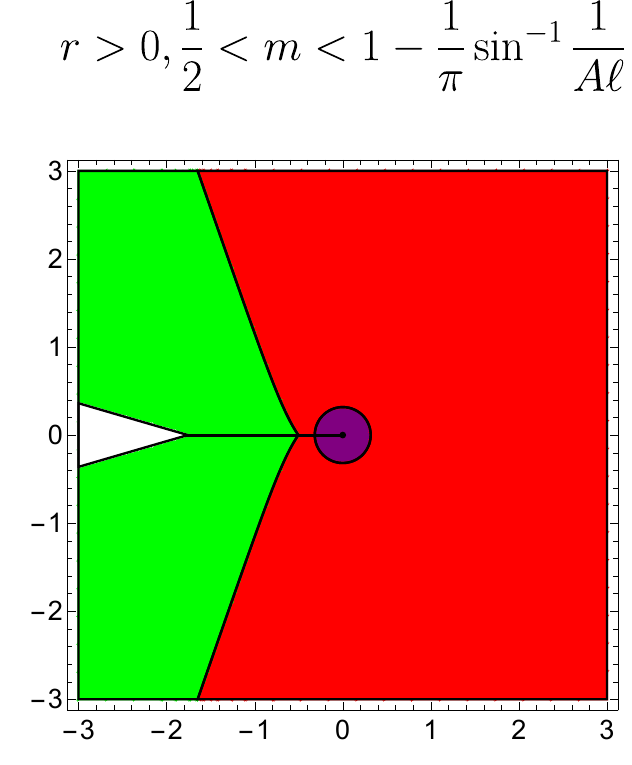}
  \caption{}
\end{subfigure}
\hfill
\begin{subfigure}{0.47\textwidth}
  \centering
  \includegraphics[scale=0.45]{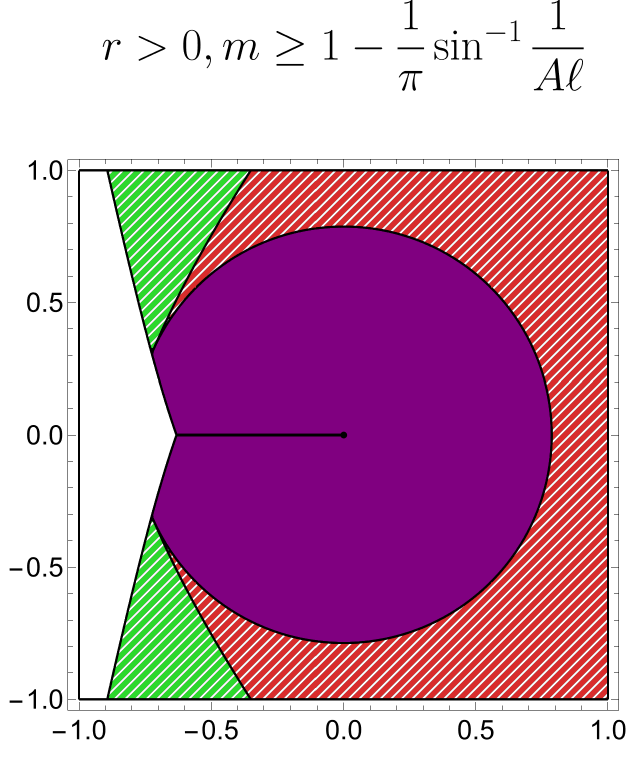}
  \caption{}
\end{subfigure}

\caption{Polar representations of the $r>0$ chart of the Class~I$_\mathrm{rapid}$ wall branch. Purple denotes the branch $s=+1$, for which the Killing coordinate $\tau$ is timelike. Light green denotes $s=-1$ with $X<1$ and spacelike $\tau$. Red denotes a region whose metric would be Euclidean if analytically extended from the $s=-1$ region, with   $X<1$. White denotes regions beyond the conformal boundary. Hatched regions are excluded from the wall/strut interpretation because they are not causally connected to both identification surfaces, $\theta=0$ and $\theta=\pi$.}
\label{fig:class-i-rapid-wall-rpos}
\end{figure}

The $r>0$ polar charts for the rapid wall branch are shown in Fig.~\ref{fig:class-i-rapid-wall-rpos}. Since the Slightly Rapid/Very Rapid distinction affects only the $r<0$ chart, this figure applies to both rapid regimes. Hatched shading indicates a geometrically present region that is excluded from the static wall analysis because it does not contain the relevant identification surface.

\paragraph{$r<0$ patch of \textit{Slightly Rapid}}

\begin{figure}[H]
\centering
\mbox{
(a)
\includegraphics[scale=0.45]{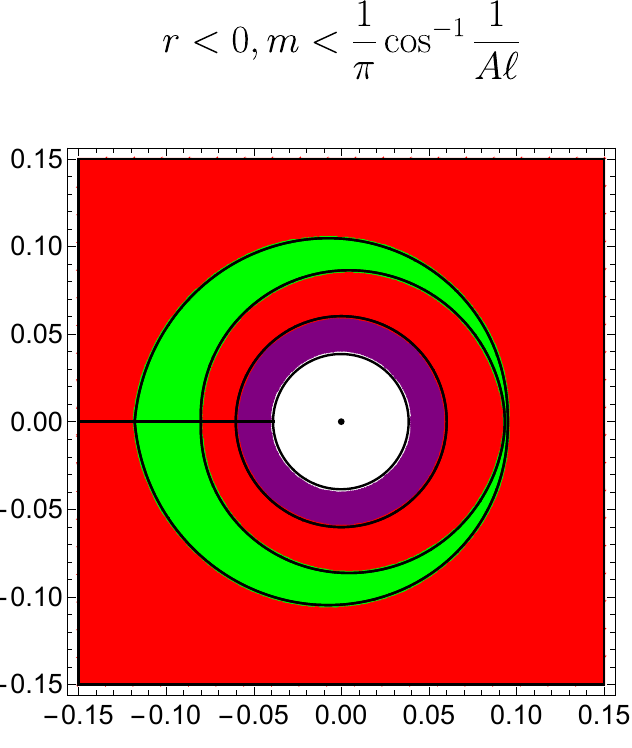}
(b)
\includegraphics[scale=0.45]{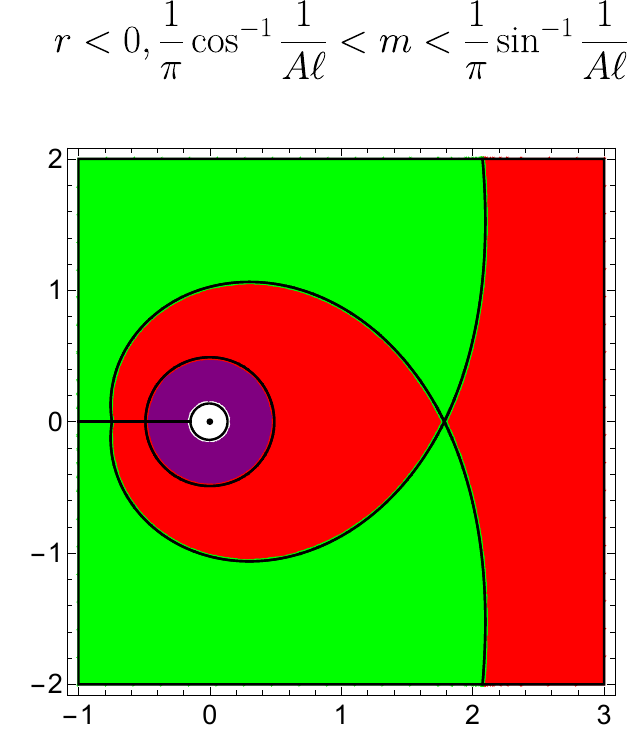}
(c)
\includegraphics[scale=0.45]{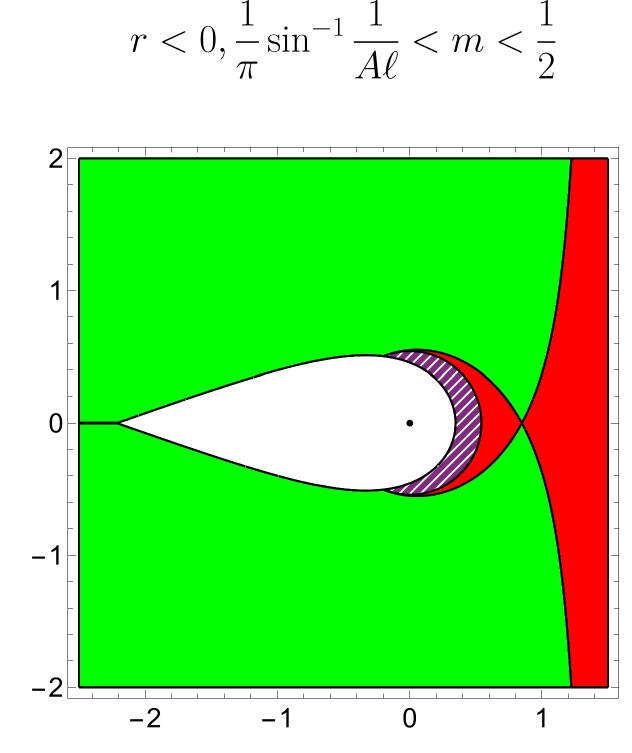}
}
\caption{The $r<0$ chart of the Slightly Rapid Class~I wall branch. (a) $0<m<\frac{1}{\pi}\cos^{-1}\frac{1}{A\ell}$. (b) $\frac{1}{\pi}\cos^{-1}\frac{1}{A\ell}<m<\frac{1}{\pi}\sin^{-1}\frac{1}{A\ell}$. (c) $\frac{1}{\pi}\sin^{-1}\frac{1}{A\ell}<m<\frac{1}{2}$.  Purple denotes the branch $s=+1$, for which the Killing coordinate $\tau$ is timelike. Light green denotes $s=-1$ with $X<1$ and spacelike $\tau$. Red denotes a region whose metric would be Euclidean if analytically extended from the $s=-1$ region, with   $X<1$. White denotes regions beyond the conformal boundary. Hatched regions are excluded from the wall/strut interpretation because they are not causally connected to both identification surfaces, $\theta=0$ and $\theta=\pi$.
}\label{fig:class-i-rapid-wall-slight-rneg}
\end{figure}

In the Slightly Rapid regime, $1<A\ell<\sqrt{2}$, the ordering $x_{X=1}>x_h$ gives three qualitatively distinct intervals of $m$ in the $r<0$ chart:
\begin{equation}\label{intervals slightly rapid}
    \left(0,\frac{1}{\pi}\cos^{-1}\!\frac{1}{A\ell}\right),\qquad
    \left(\frac{1}{\pi}\cos^{-1}\!\frac{1}{A\ell},\frac{1}{\pi}\sin^{-1}\!\frac{1}{A\ell}\right),\qquad
    \left(\frac{1}{\pi}\sin^{-1}\!\frac{1}{A\ell},\frac{1}{2}\right).
\end{equation}
Fig.~\ref{fig:class-i-rapid-wall-slight-rneg} displays these three cases from left to right. They correspond to the three possible relative positions of the conformal boundary, the endpoint $f(r)=0$, and the curve $X=1$ within the $r<0$ branch. The first and third panels are qualitatively equivalent to the outer intervals of the Very Rapid case, while the middle panel is specific to the Slightly Rapid ordering.

\paragraph{$r<0$ patch of \textit{Very Rapid}}

\begin{figure}[H]
\centering
\mbox{
(a)
\includegraphics[scale=0.45]{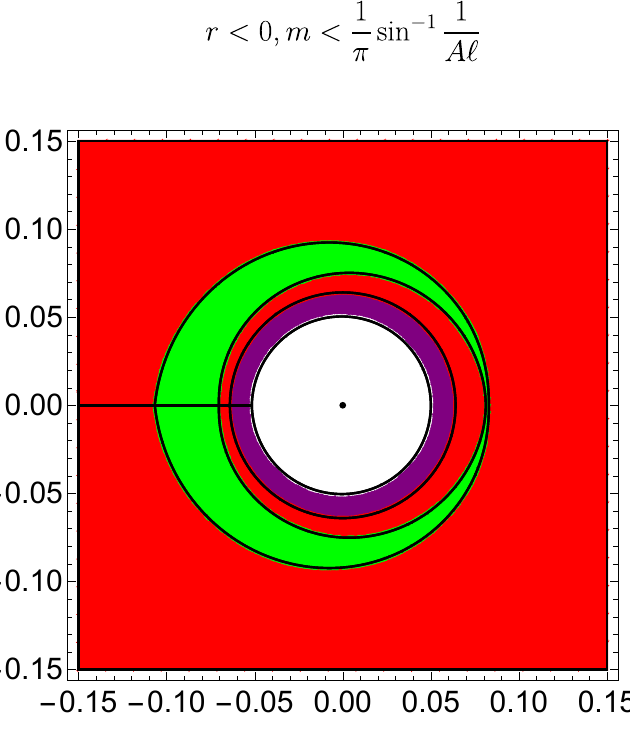}
(b)
\includegraphics[scale=0.45]{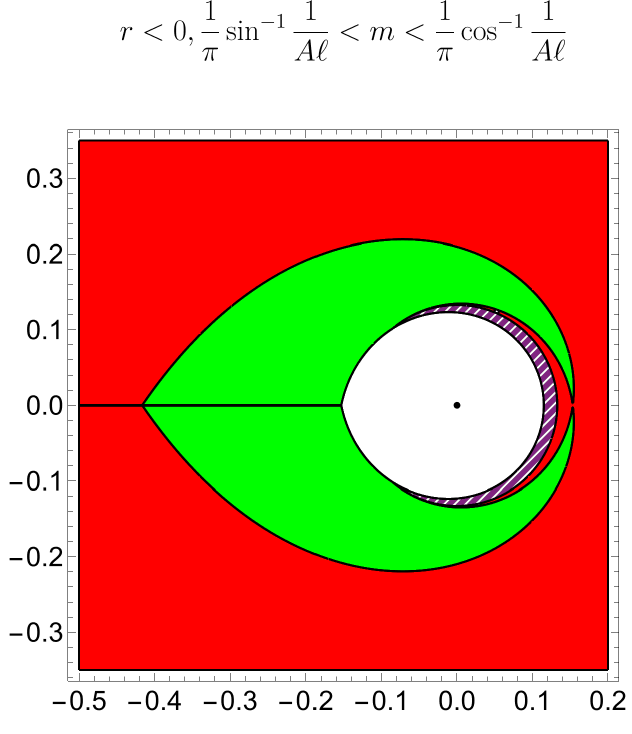}
(c)
\includegraphics[scale=0.45]{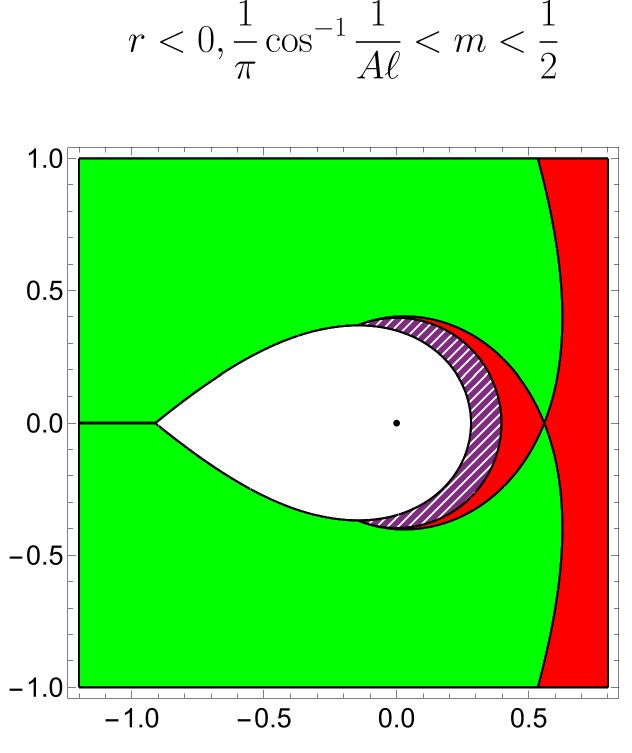}
}
\caption{The $r<0$ chart of the Very Rapid Class~I wall branch: (a) $0<m<\frac{1}{\pi}\sin^{-1}\frac{1}{A\ell}$; \\ (b) $\frac{1}{\pi}\sin^{-1}\frac{1}{A\ell}<m<\frac{1}{\pi}\cos^{-1}\frac{1}{A\ell}$;  (c) $\frac{1}{\pi}\cos^{-1}\frac{1}{A\ell}<m<\frac{1}{2}$. Purple denotes the branch $s=+1$, for which the Killing coordinate $\tau$ is timelike. Light green denotes $s=-1$ with $X<1$ and spacelike $\tau$. Red denotes a region whose metric would be Euclidean if analytically extended from the $s=-1$ region, with   $X<1$. White denotes regions beyond the conformal boundary. Hatched regions are excluded from the wall/strut interpretation because they are not causally connected to both identification surfaces, $\theta=0$ and $\theta=\pi$.
}
\label{fig:class-i-rapid-wall-very-rneg}
\end{figure}

In the Very Rapid regime, $A\ell>\sqrt{2}$, one has $x_{X=1}<x_h$, so the ordering of the relevant curves in the $r<0$ chart is reversed relative to the Slightly Rapid case. The three intervals of $m$ are
\begin{equation}\label{intervals very rapid}
    \left(0,\frac{1}{\pi}\sin^{-1}\!\frac{1}{A\ell}\right),\qquad
    \left(\frac{1}{\pi}\sin^{-1}\!\frac{1}{A\ell},\frac{1}{\pi}\cos^{-1}\!\frac{1}{A\ell}\right),\qquad
    \left(\frac{1}{\pi}\cos^{-1}\!\frac{1}{A\ell},\frac{1}{2}\right).
\end{equation}
Fig.~\ref{fig:class-i-rapid-wall-very-rneg} displays these three intervals from left to right. The first and third intervals are qualitatively equivalent to the corresponding outer intervals of Fig.~\ref{fig:class-i-rapid-wall-slight-rneg}; only the middle interval is unique to the Very Rapid ordering. For this reason, we retain the Slightly Rapid and Very Rapid terminology rather than rewriting the two classifications in a unified min/max notation.

\begin{figure}[H]
\centering
\mbox{
(a)
\includegraphics[scale=0.45]{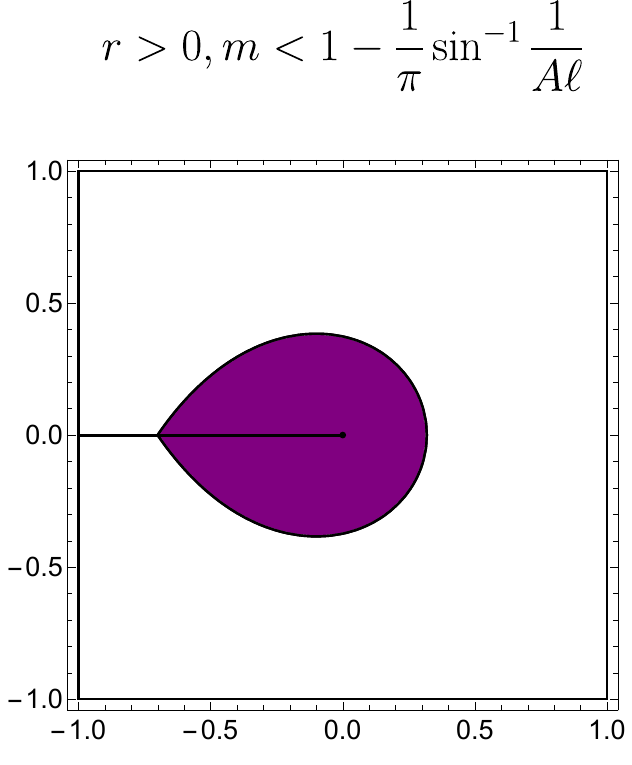}
(b)
\includegraphics[scale=0.45]{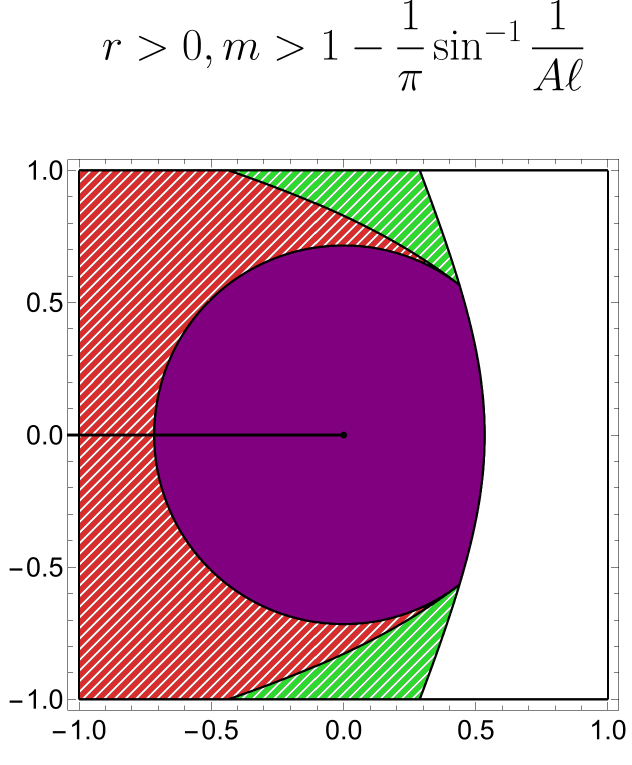}
}
\caption{Polar representations of the Class~I$_\mathrm{rapid}$ strut branch. Purple denotes the branch $s=+1$, for which the Killing coordinate $\tau$ is timelike. Purple denotes the branch $s=+1$, for which the Killing coordinate $\tau$ is timelike. Light green denotes $s=-1$ with $X<1$ and spacelike $\tau$. Red denotes a region whose metric would be Euclidean if analytically extended from the $s=-1$ region, with   $X<1$. White denotes regions beyond the conformal boundary. Hatched regions are excluded from the wall/strut interpretation because they are not causally connected to both identification surfaces, $\theta=0$ and $\theta=\pi$}.
\label{fig:class-i-rapid-strut-polar}
\end{figure}

\paragraph{Strut}

For the rapid strut branch, the relevant polar domains are shown in Fig.~\ref{fig:class-i-rapid-strut-polar}. The $r<0$ chart is irrelevant to the present analysis because only the lower purple component, corresponding to $s=+1$ with timelike $\tau$, is connected to the regular identification surface at $x=-1$ required for the static strut construction. We therefore discard the $r<0$ chart together with the $s=-1$ sectors. Since the distinction between the Slightly Rapid and Very Rapid regimes affects only the discarded $r<0$ chart, no additional subdivision is needed for the rapid strut case.

\subsection{Class II}

In the C-metric, Class~II is the accelerating-black-hole branch with $Q(x)=x^2-1$ 
\cite{ArenasHenriquez:2023hur}. 
For the solutions we consider,
the two disconnected prolate ranges, $x>1$ and $x<-1$, lead to the right and left branches discussed below.

\paragraph{Right}

\begin{figure}[H]
    \centering
    \includegraphics[scale=0.45]{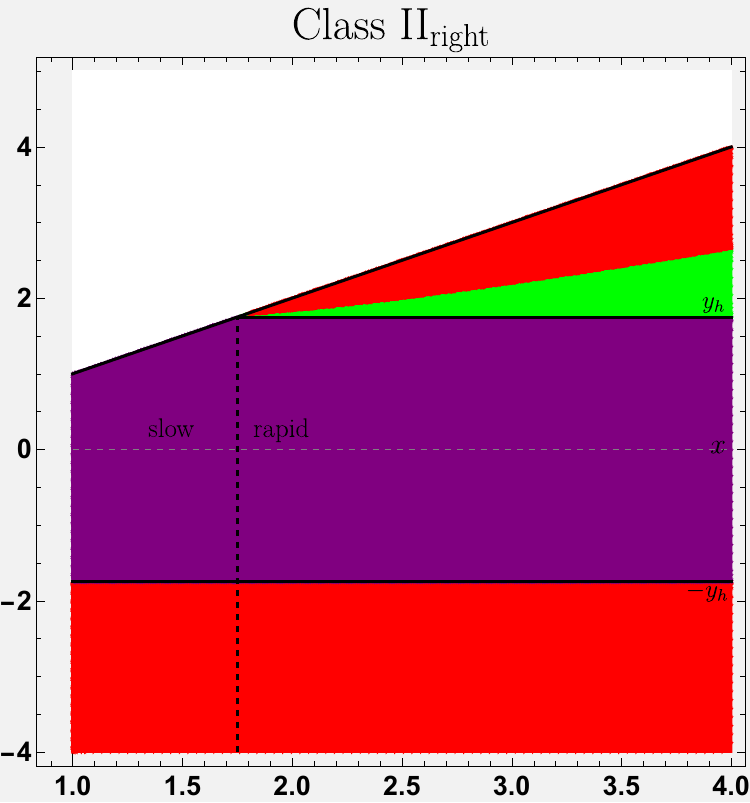}
    \caption{Coordinate domain of Class~II$_\mathrm{right}$ in the $(x,y)$ plane. The dashed vertical line separates the slow regime, $x_+\le y_h$, from the rapid regime, $x_+>y_h$. The upper triangular sector is not used in the connected polar construction. Purple denotes the branch $s=+1$, for which the Killing coordinate $\tau$ is timelike. Light green denotes $s=-1$ with $X<1$ and spacelike $\tau$. Red denotes a region whose metric would be Euclidean if analytically extended from the $s=-1$ region, with   $X<1$. White denotes regions beyond the conformal boundary.}
    \label{fig:class-ii-right-prolate}
\end{figure}

\begin{figure}[H]
\centering

\begin{subfigure}{0.47\textwidth}
  \centering
  \includegraphics[scale=0.45]{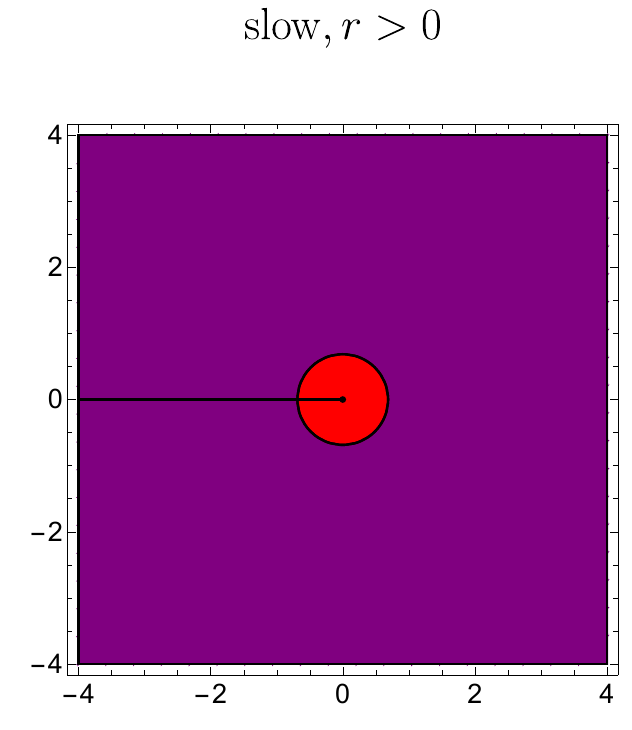}
  \caption{Slow, $r>0$.}
\end{subfigure}
\hfill
\begin{subfigure}{0.47\textwidth}
  \centering
  \includegraphics[scale=0.45]{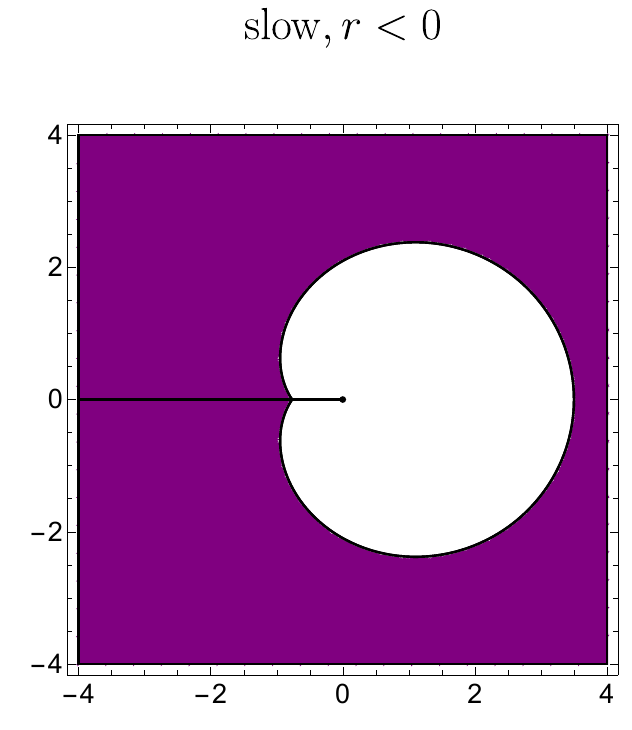}
  \caption{Slow, $r<0$.}
\end{subfigure}

\vspace{0.4cm}
\begin{subfigure}{0.47\textwidth}
  \centering
  \includegraphics[scale=0.45]{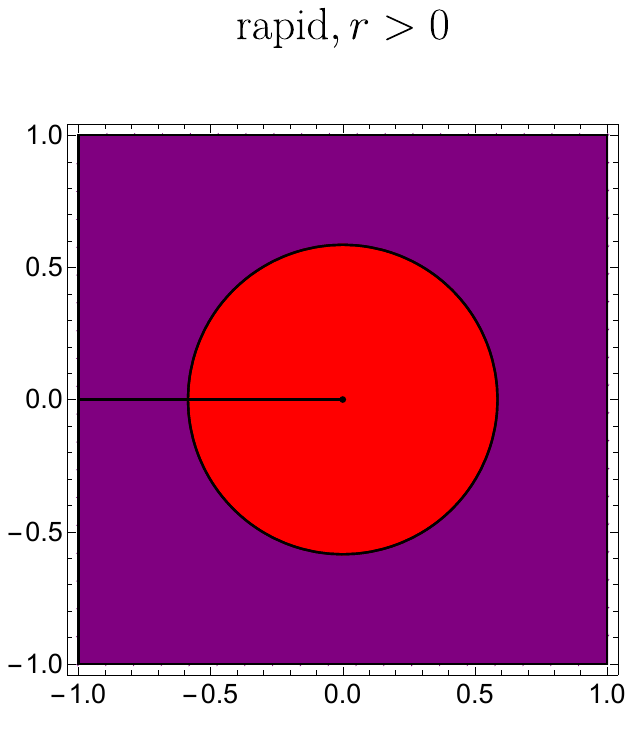}
  \caption{Rapid, $r>0$.}
\end{subfigure}
\hfill
\begin{subfigure}{0.47\textwidth}
  \centering
  \includegraphics[scale=0.45]{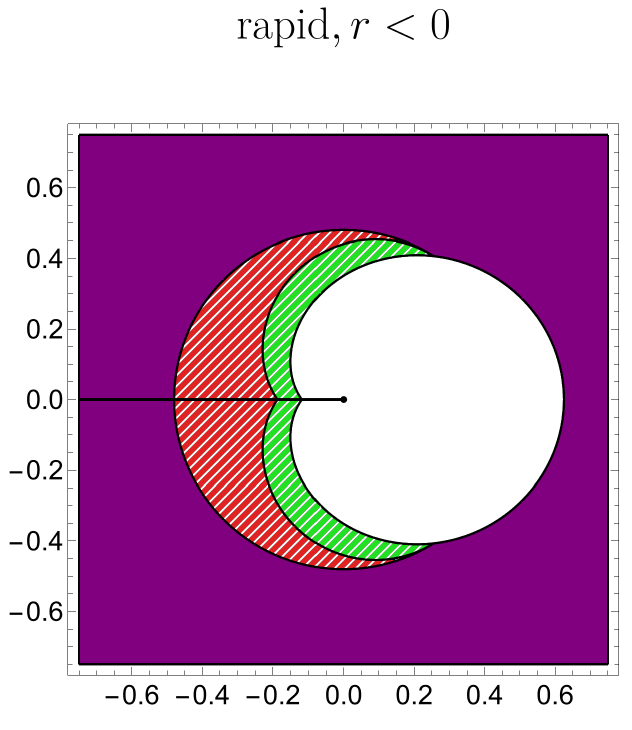}
  \caption{Rapid, $r<0$.}
\end{subfigure}
\hfill

\caption{Polar representations of Class~II$_\mathrm{right}$.Purple denotes the branch $s=+1$, for which the Killing coordinate $\tau$ is timelike. Light green denotes $s=-1$ with $X<1$ and spacelike $\tau$. Red denotes a region whose metric would be Euclidean if analytically extended from the $s=-1$ region, with   $X<1$. White denotes regions beyond the conformal boundary. Hatched regions are excluded from the wall/strut interpretation because they are not causally connected to both identification surfaces, $\theta=0$ and $\theta=\pi$.}
\label{fig:class-ii-right-polar}
\end{figure}

Class~II$_\mathrm{right}$ corresponds to the range $x>1$. In prolate coordinates, it has two regimes, shown in Fig.~\ref{fig:class-ii-right-prolate}: the slow case $x_+\leq y_h$ and the rapid case $x_+>y_h$. In the slow case, the marker $f(r)=0$ is excluded from the $y>0$ ($r<0$) patch, whereas in the rapid case, it is included as the endpoint of the spatial slice. The upper triangular sector in Fig.~\ref{fig:class-ii-right-prolate} does not contain the identification surface required for the connected static construction, so it is omitted from the polar-domain plots in Fig.~\ref{fig:class-ii-right-polar}. Both the $r>0$ and $r<0$ charts exist for this branch.

Fig.~\ref{fig:class-ii-right-polar} gives the polar representation of the same right branch. The $r>0$ chart is qualitatively the same in the slow and rapid regimes. The difference appears in the $r<0$ chart: in the slow case, the endpoint $f(r)=0$ lies outside the physical domain, whereas in the rapid case, it lies within the displayed domain. Hatched regions are shown only to indicate excluded pieces that do not contribute to the connected static strut interpretation.

\paragraph{Left}

\begin{figure}[H]
    \centering
    \includegraphics[scale=0.45]{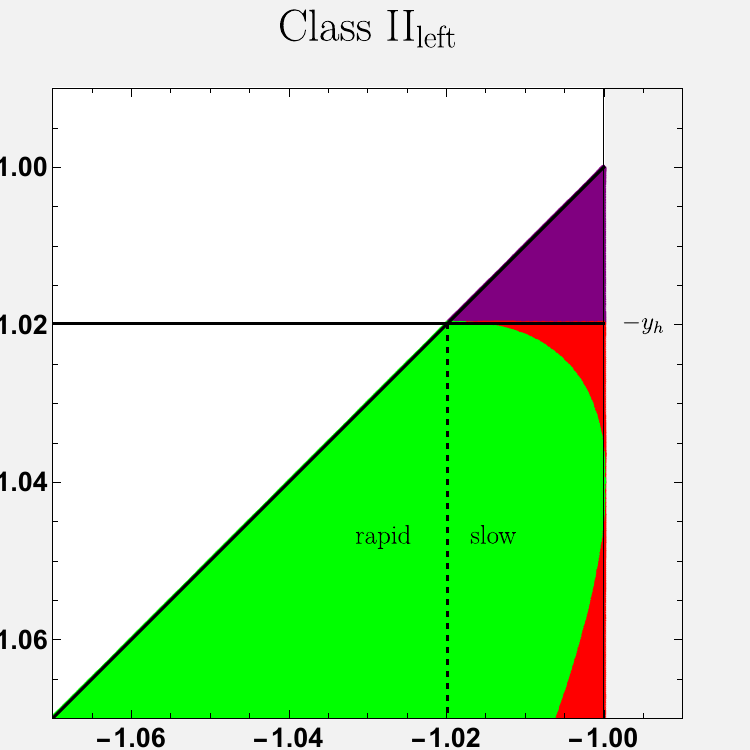}
    \caption{Coordinate domain of Class~II$_\mathrm{left}$ in the $(x,y)$ plane. The slow and rapid regimes are separated by the condition $x_+=-y_h$. Purple denotes the branch $s=+1$, for which the Killing coordinate $\tau$ is timelike. Light green denotes $s=-1$ with $X<1$ and spacelike $\tau$. Red denotes a region whose metric would be Euclidean if analytically extended from the $s=-1$, with the $X<1$ case. White denotes regions beyond the conformal boundary.}
    \label{fig:class-ii-left-prolate}
\end{figure}

\begin{figure}[H]
\centering
\mbox{
(a)
\includegraphics[scale=0.45]{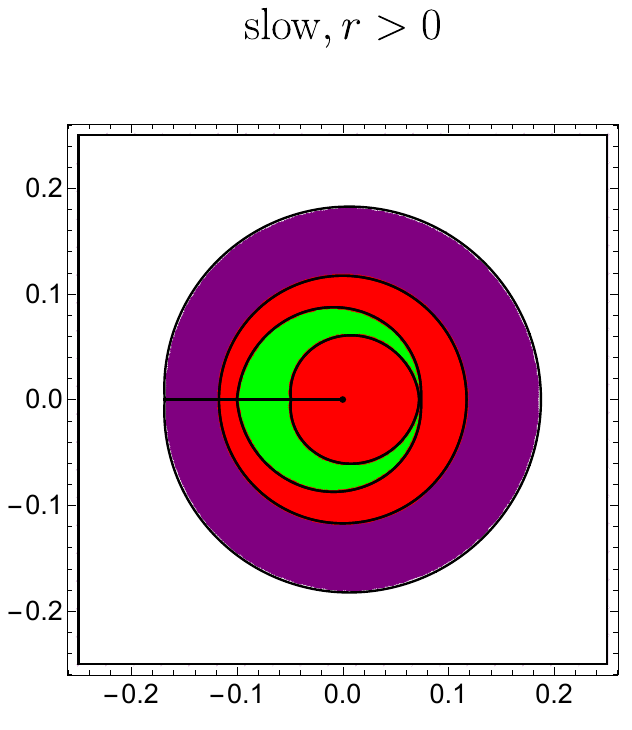}
(b)
\includegraphics[scale=0.45]{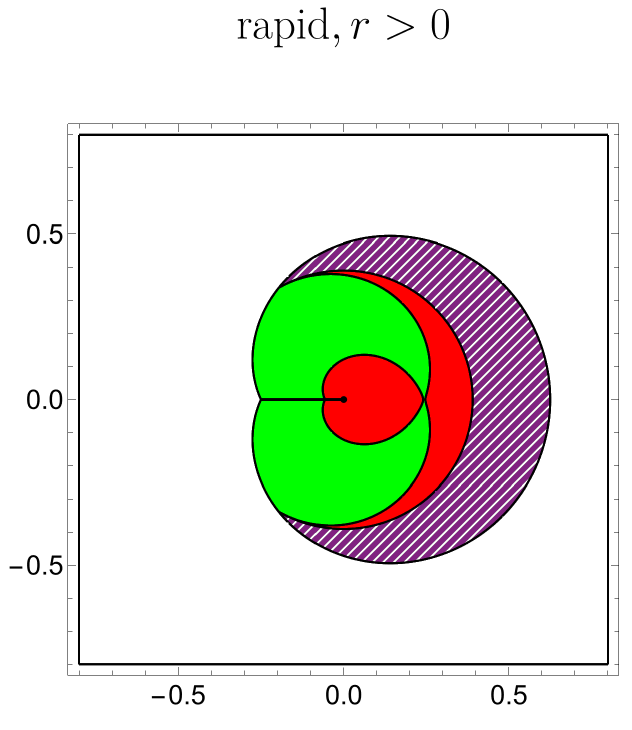}
}
\caption{Polar representations of the Class~II$_\mathrm{left}$ wall branch: (a) Slow; (b) Rapid. Purple denotes the branch $s=+1$, for which the Killing coordinate $\tau$ is timelike. Light green denotes $s=-1$ with $X<1$ and spacelike $\tau$. Red denotes a region whose metric would be Euclidean if analytically extended from the $s=-1$ region, with   $X<1$. White denotes regions beyond the conformal boundary. Hatched regions are excluded from the wall/strut interpretation because they are not causally connected to both identification surfaces, $\theta=0$ and $\theta=\pi$.}
\label{fig:class-ii-left-polar}
\end{figure}

Class~II$_\mathrm{left}$ corresponds to the range $x<-1$. As shown in Fig.~\ref{fig:class-ii-left-prolate}, it also has slow and rapid regimes, depending on whether the identification satisfies $x_+\ge -y_h$ or $x_+<-y_h$. Unlike Class~II$_\mathrm{right}$, only the $y<0$ ($r>0$) chart is relevant for the single-defect construction. The polar plots in Fig.~\ref{fig:class-ii-left-polar} show that, in the rapid case, the defect locus at $\theta=\pm\pi$ lies outside the relevant timelike $s=+1$ sector; that sector is therefore hatched and excluded. In the slow case, the defect remains within the relevant static branch.

\subsection{Class III}

\begin{figure}[H]
    \centering
    \includegraphics[scale=0.45]{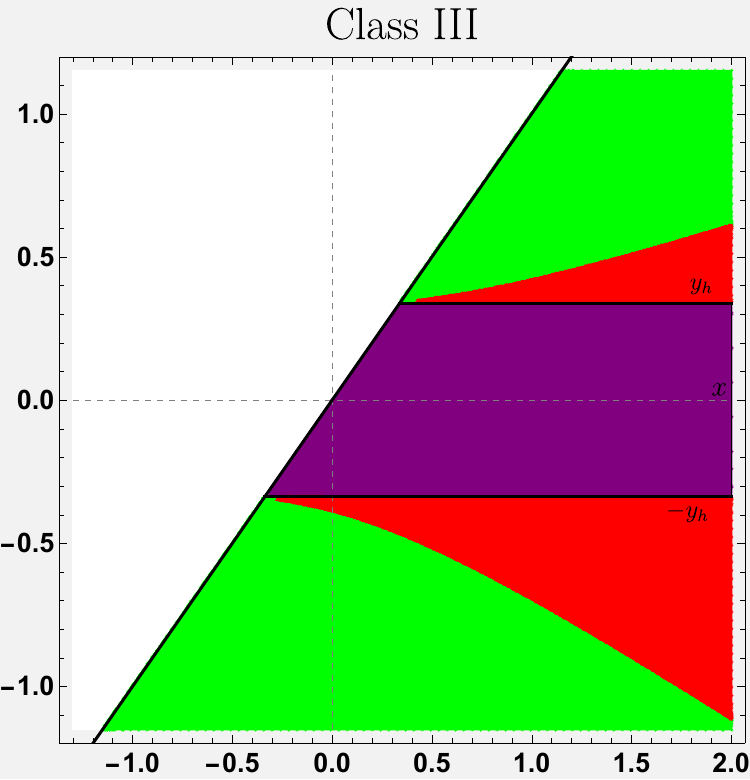}
    \caption{Coordinate domain of the Class~III branch for $A\ell<1$. Purple denotes the branch $s=+1$, for which the Killing coordinate $\tau$ is timelike. Light green denotes $s=-1$ with $X<1$ and spacelike $\tau$. Red denotes a region whose metric would be Euclidean if analytically extended from the $s=-1$ region, with $X<1$. White denotes regions beyond the conformal boundary.}
    \label{fig:class-iii-prolate}
\end{figure}

Class~III is shown in Fig.~\ref{fig:class-iii-prolate}. We do not analyze it in the same detail as Classes~I and II, since it is more naturally interpreted as a braneworld-type construction than as a single-defect particle or black-hole geometry. There are several qualitatively distinct possibilities depending on the value of $A\ell$. If $A\ell>1$, the spacetime is everywhere spacelike with respect to $\tau$; if $A\ell=1$, it is spacelike everywhere and becomes null only at $y_h=0$. We therefore show only the case $A\ell<1$, where the original timelike region is replaced by disconnected $s=-1$, $X<1$ and $s=-1$, $X>1$ branches. Any identification that yields a black hole or particle interpretation would require cutting and gluing two copies of the spacetime along two independent surfaces, introducing two defects rather than a single fundamental wall or strut. For this reason, we do not present an identified polar version of Class~III here.

\section{The domain wall}\label{sec:pro}

We now consider the behavior of the metric at $\theta = \pm \pi$, introducing a domain wall to ensure the field equations hold, analogous to the situation for the C-metric
\cite{Astorino_2011,Arenas_Henriquez_2022}.   

We note that $\mathcal{E}_{ab} = 0$  for any $|\theta|\not=\pi$. We also note that due to the mirror symmetry, any scalar should be an even function around $\theta=\pm\pi$. Henceforth, we denote the side where
$\theta > -\pi$  with a minus sign $-$ and the 
side where $\theta < \pi$
 with a plus sign $+$. We use a square bracket to represent the difference between boundary terms on these sides:
\begin{eqnarray}
    [T]=T^--T^+.
\end{eqnarray}

Around the junction, we have the decomposition 
\begin{eqnarray}\label{Rsplit}
    R_{ab}&=&\Theta(l)R^-_{ab}+\Theta(-l)R^+_{ab}+\delta(l)A_{ab},\\
    R&=&R_\pi+\delta(l)A,
\end{eqnarray}
where $\Theta$ is the Heaviside function, 
$dl=n^{a}dx_a$  and $n^{\alpha}$ is the normal vector of the domain wall (string) hypersurface.  Here 
\begin{eqnarray}
    A_{ab}=\frac{1}{2} \left(\kappa_{c a} n^c n_b+\kappa_{c b} n^c n_a-\kappa n_a n_b -\kappa_{ab} \right),
\end{eqnarray}
where
\begin{eqnarray}
    \kappa_{ab} \equiv [g_{ab,c}]n^c.\label{kappadef}
\end{eqnarray}
Note here that $R$ is not partitioned into positive and negative parts because it is a scalar and therefore has the same value in the neighborhood of the boundary with mirror symmetry.

The scalar field decomposes as  
\begin{eqnarray}
\nabla_a\nabla_b\phi&=&\Theta(l)\nabla_a\nabla_b\phi^-+\Theta(-l)\nabla_a\nabla_b\phi^+ + \delta(l)n_a[\nabla_b\phi],\label{eq47}\\
\Box\phi&=&\Box\phi_\pi+\delta(l)n^e[\nabla_e\phi].\label{eq48}
\end{eqnarray}
Note that since $\nabla_a\nabla_b\phi=\nabla_b\nabla_a\phi$ we obtain 
\begin{eqnarray}
n_a[\nabla_b\phi]=n_b[\nabla_a\phi]
\end{eqnarray}
from Eq.~\eqref{eq47}. 

We find that Eq.~\eqref{zero-field} is smooth across the domain wall, as shown in the Appendix.
After some computation we find that $\mathcal{E}_{ab}$ decomposes as 
\begin{eqnarray} \mathcal{E}_{ab}=\Theta(-l)\mathcal{E}_{ab}^{+(0)}+\Theta(l)\mathcal{E}_{ab}^{-(0)}+\mathcal{E}_{ab}^{(1)}\delta(l),
\end{eqnarray}
where $\mathcal{E}_{ab}^{+(0)}=\mathcal{E}_{ab}^{-(0)} = 0$  due to Eq.~\eqref{Eabsimp}. 
 Therefore, the junction term, which is proportional to $\delta(l)$, is the physical contribution that we need.
Finally, we obtain
\begin{eqnarray}
    \mathcal{E}^{(1)}_{\tau\tau}&=&\left(A_{\tau\tau}-\frac{1}{2}Ag_{\tau\tau}\right)(1+2\alpha(\nabla\phi)^2)+4\alpha g_{\tau\tau
    } n^x[\nabla_x\phi]\Big(\nabla^y\nabla_y\phi- \nabla^{x} \phi\nabla_{x} \phi\Big),\label{Ett1}\\
    \mathcal{E}^{(1)}_{yy}&=&\left(A_{yy}-\frac{1}{2}Ag_{yy}\right)(1+2\alpha(\nabla\phi)^2)+4\alpha g_{yy}n^x[\nabla_x\phi]\Big(\nabla^\tau\nabla_\tau\phi- \nabla^x \phi\nabla_x \phi-\nabla^y\phi\nabla_y\phi\Big).\label{Err1}
\end{eqnarray}

We find that it is more convenient to work in polar coordinates when calculating the junction conditions; hence, we obtain  
\begin{eqnarray}
    8\pi\mathcal{E}^{(1)\tau}{}_{\tau}&=&\frac{1}{2}\kappa^r{}_{r}\left(1+2\alpha(\nabla\phi)^2\right)
    +2\alpha  \kappa^\tau{}_\tau\Big(\nabla^r\nabla_r\phi- \nabla^{\theta} \phi\nabla_{\theta} \phi\Big),\\
    8\pi\mathcal{E}^{(1)r}{}_{r}&=&\frac{1}{2}\kappa^\tau{}_{\tau}\left(1+2\alpha(\nabla\phi)^2\right).
\end{eqnarray}

As we will only be calculating the tension along the cosmic strut/string, we will frequently use $X_\pi$, which is the value of $X$ when $\theta=\pi$ (i.e., along the $r$-axis):
\begin{equation}
    X_\pi\equiv X_\pi(r) \equiv \left|A\ell m\tilde{\Omega}(r,\pi)\sqrt{\frac{P(0)}{f(r)}}\right|=
    \begin{cases}
        |m+Ar \cos(m\pi)|\left|\frac{r^2}{\ell^2}+\frac{m}{1-A^2\ell^2}\right|^{-1/2}, &\textbf{Class I}_\mathrm{wall},
        \\|m-Ar \cos(m\pi)|\left|\frac{r^2}{\ell^2}+\frac{m}{1-A^2\ell^2}\right|^{-1/2}, &\textbf{Class I}_\mathrm{strut},
        \\|m+Ar \cosh(m\pi)|\left|\frac{r^2}{\ell^2}-\frac{m}{1+A^2\ell^2}\right|^{-1/2}, &\textbf{Class II}_\mathrm{right},
        \\|m-Ar \cosh(m\pi)|\left|\frac{r^2}{\ell^2}-\frac{m}{1+A^2\ell^2}\right|^{-1/2}, &\textbf{Class II}_\mathrm{left}.
    \end{cases}
\end{equation}
We interpret $- \mathcal{E}^{(1)\tau}{}_{\tau} \equiv \sigma$ as the tension and $\mathcal{E}^{(1)r}{}_{r} \equiv \lambda$ as the pressure of the wall. 

We are only interested in defects that admit a static string or strut interpretation. Thus, we keep only those components for which the identification surface at $\theta=\pm\pi$ is timelike with respect to the Killing coordinate $\tau$. In Class~I$_\mathrm{slow}$, the whole physical domain lies in the branch $s=-1$, $X<1$, and this branch has timelike $\tau$; hence, the wall calculation below is carried out with $s=-1$ for this case. In Class~I$_\mathrm{rapid}$, the only branch relevant for a static wall or strut is the $s=+1$ branch. The $s=-1$ branch is spacelike with respect to $\tau$, and an identification there would describe a non-static defect, so we do not include it in the tension calculation.

For Class~II, we likewise compute the tension only on the $s=+1$ branch. The remaining $s=-1$ sector is spacelike with respect to $\tau$ and therefore does not give a static string or strut.  In the   C-metric in (2+1)-dimensional Einstein gravity \cite{Astorino_2011,Arenas_Henriquez_2022}, Class~III is better interpreted as a braneworld-type construction rather than as a single-defect particle or black-hole geometry.
We therefore do not make a $\theta=\pm\pi$ identification for Class~III, and we do not assign a string tension to it in the following analysis. With these restrictions, the tension and equation of state $\sigma,w$ along the wall are \cite{Musgrave_1996}
\begin{align}\label{EoS}
\sigma&=\frac{ A \sqrt{B_\alpha } \sqrt{\frac{4 \alpha }{l^2}+1}}{4\pi\sqrt{1+sX_{\pi} ^2} } \left(1+s\frac{X_{\pi}  H(X_{\pi} )}{B_\alpha }\right)\mathrm{sgn}\left(H(X_{\pi} )\right)
\times \begin{cases}
    \sin (m\pi),&\textbf{Class I}_\mathrm{wall},\\
    -\sin (m\pi),&\textbf{Class I}_\mathrm{strut},\\
    (-\sinh (m\pi)),&\textbf{Class II}_\mathrm{right},\\
    \sinh (m\pi),&\textbf{Class II}_\mathrm{left},
\end{cases}\\
    w& \equiv \frac{\lambda}{\sigma} =-\left(1+s \frac{ X_\pi H(X_\pi)}{B_\alpha}\right)^{-1}
    =\begin{cases} 
          -\left({1- X_\pi\left(\frac{H_0}{B_\alpha}+ \tanh^{-1} X_\pi\right)}\right)^{-1}, & s=-1, X_\pi<1, \\
          -\left({1+ X_\pi\left(\frac{H_0}{B_\alpha}+ \tan^{-1} X_\pi\right)}\right)^{-1},   & s=+1.
       \end{cases}
\end{align}

Setting $H_0=0$, we have
\begin{align}\label{wH0}
    w& \equiv \frac{\lambda}{\sigma} 
    =\begin{cases} 
          -\left({1- X_\pi\tanh^{-1} X_\pi}\right)^{-1}, & s=-1, X_\pi<1, \\
          -\left({1+ X_\pi\tan^{-1} X_\pi}\right)^{-1},   & s=+1,
       \end{cases}
\end{align}
where the null energy condition (NEC) requires 
$\sigma > 0$ and $w \geq -1$. It is interesting to note that for the standard C-metric cases, the equation of state is always $w = -1$
(and hence, a constant tension), while the tension signature can be either positive or negative.

For the choice $H_0=0$, the $s=+1$ branch always satisfies $w\geq -1$, since $X_\pi\tan^{-1}X_\pi\geq0$. Therefore, on the static $s=+1$ branches of Class~I$_\mathrm{rapid}$ and Class~II, the NEC reduces to the sign condition $\sigma>0$. By contrast, in the $s=-1$, $X_\pi<1$ branch relevant for Class~I$_\mathrm{slow}$, one has
\begin{equation}
w=-\left(1-X_\pi\tanh^{-1}X_\pi\right)^{-1}.
\end{equation}
Thus $w<-1$ whenever $0<X_\pi\tanh^{-1}X_\pi<1$. Equivalently, the NEC is violated for $0<X_\pi<X_\star$, where $X_\star$ is defined by
\begin{equation}
X_\star\tanh^{-1}X_\star=1,
\qquad
X_\star\simeq0.833557.
\end{equation}
For $X_\pi>X_\star$, the equation-of-state part of the NEC is satisfied, again leaving the sign of $\sigma$ as the remaining condition.

Especially, in I$_\mathrm{Saturated}$,
\begin{align}
\sigma&=\frac{A}{4\pi} \sqrt{B_\alpha } \sqrt{\frac{4 \alpha }{l^2}+1} \mathrm{sgn}\left(H(X_{\pi} )\right)
\times \begin{cases}
    \sin (m\pi),&\textbf{Class I}_\mathrm{wall},\\
    -\sin (m\pi),&\textbf{Class I}_\mathrm{strut},\\
\end{cases}\\
w&=-1.
\end{align}
$w$ is equal to that of the unmodified C-metric, which means that the NEC is always fulfilled in I$_\mathrm{Saturated}$, while the tension is equal to the tension of the unmodified case in the $\alpha\to 0$ limit. 

\section{\label{sec:dis}Discussion}

We have obtained several new classes of solutions, Eq.~\eqref{C1A} and Eq.~\eqref{C23A}, in 3DEGB that generalize the C-metric in 3D Einstein gravity. In so doing, we have obtained new locally AdS metrics that, to our knowledge, have not appeared in the literature.
These classes of solutions we have found contain a number of free parameters, each with markedly different behavior.   

For $\alpha\neq 0$ we obtain a previously unknown representation of AdS spacetime that is also a solution to the $(2+1)$-dimensional Einstein equations but with the cosmological constant rescaled as in Eq.~\eqref{Rscal}. Curiously, the $\alpha\to 0$ limit does not recover the standard C-metric \cite{Arenas_Henriquez_2022}, and the scalar field does not change the spacetime in any significant manner other than by rescaling the AdS length-scale. 

 Our solutions require the existence of a thin string (domain wall). The physical meaning is that our solutions are accelerating objects, with the acceleration provided by the physical object (a thin string). 
They do not correspond to either black holes or point particles, but rather to 
solitons, or alternatively spacetimes containing ``bubbles of nothing". 
 Although we have concentrated only on subclass A in this paper, this feature is generically true for all of the subclasses we have found.

 Several questions arise. Is there any explicit connection to our solution for accelerated geometry, such as a Rindler space? How do  quantum effects modify the  accelerating geometry? What is the interpretation of solutions for which $f(r)<0$, which to non-static spacetimes? These topics have  been   investigated in  Einstein gravity case to varying extents, but they are not well understood  in 3DEGB.  
 
 In addition, we need more analysis of Classes II and III, as well as Subclasses B-F. They are not accelerating black holes but rather something else. We also leave these topics for future investigation.

\section*{Acknowledgements}
This work was supported in part by the Natural Sciences and Engineering Research Council of Canada.  We are grateful to Ruth Gregory for helpful discussions. DY was supported by the National Research Foundation of Korea (NRF) grant funded by the Korean government (No. RS-2026-25476711).

\newpage

\appendix

\section{Embedding of the metric}
 In this section, we use the polar coordinate conventions summarized in Table~\ref{table::ClassesPolar}.

\subsection{Poincar\'e patch}

As we have established that the spacetime is locally AdS,  we can express the metric in Poincar\'e patch coordinates \((T,Z,L)\). We obtain
\begin{align}
    ds^2&=\frac{\ell^2}{B_\alpha Z^2}(-dT^2+dZ^2+dL^2),\\
    T(\tau)&=\sqrt{|B_\alpha P(0)|}\tau,\\
    Z(r,\theta)&=H(X(r,\theta)),\\
    L(r,\theta)&=\begin{cases}
        B_\alpha \cot ^{-1}\left(\sqrt{1-A^2 \ell^2} \cot (m\theta) -\frac{A \ell^2 m}{r \sqrt{1-A^2 \ell^2} }\frac{1}{\sin (m\theta)}\right), & \textbf{Class~I}, A\ell<1,
        \\B_\alpha \coth ^{-1}\left(\sqrt{A^2 \ell^2-1} \cot (m\theta)+\frac{A \ell^2 m}{r \sqrt{A^2 \ell^2-1} }\frac{1}{\sin (m\theta)}\right), & \textbf{Class~I}, f(r)>0, A\ell>1,
        \\B_\alpha\coth ^{-1}\left(\sqrt{1+A^2\ell^2}\coth(m\theta)+\frac{A \ell^2 m }{r \sqrt{1+A^2 \ell^2}}\frac{1}{\sinh(m\theta)}\right), &\textbf{Class~II}, f(r)>0,
        \\ B_\alpha \tanh ^{-1}\left(\sqrt{1-A^2 \ell^2} \tanh (m\theta)-\frac{A\ell^2m}{r\sqrt{1-A^2\ell^2}}\frac{1}{\cosh{(m\theta)}}\right), &\textbf{Class~III}, f(r)>0   , A\ell<1 ,  
    \end{cases}\label{LPoincareA}
\end{align}
and
\begin{equation}
    \phi=-\log |Z|.
\end{equation}

Note that assuming $r>0$, $L(r,\theta)$ is real in each   Class and region of Eq.~\eqref{LPoincareA}. To see this explicitly, use the standard real branches
\begin{equation}
\cot^{-1}:\mathbb{R}\to(0,\pi),\qquad
\coth^{-1}:(-\infty,-1)\cup(1,\infty)\to\mathbb{R},\qquad
\tanh^{-1}:(-1,1)\to\mathbb{R}.
\end{equation}
For $f(r)>0$ and Class~I with $A\ell<1$, the argument of $\cot^{-1}$ is
\begin{equation}
u_{I,<}=\sqrt{1-A^2\ell^2}\,\cot m\theta-\frac{A\ell^2m}{r\sqrt{1-A^2\ell^2}}\frac{1}{\sin m\theta},
\end{equation}
which is manifestly real whenever $\sin m\theta\neq0$. At $\sin m\theta=0$ $u_{I,<}$ tends to $-\infty$, so $L(r,\theta)=B_\alpha\cot^{-1}u_{I,<}$ still has a real limiting value.

For $f(r)>0$ and Class~I with $A\ell>1$, the argument of $\coth^{-1}$ is
\begin{equation}
    u_{I,>} = \sqrt{A^2 \ell^2-1} \cot (m\theta)+\frac{A \ell^2 m}{r \sqrt{A^2 \ell^2-1} }\frac{1}{\sin (m\theta)}.
\end{equation}
Since $0<r<\frac{m}{\sqrt{A^2\ell^2-1}}$ for $f(r)>0$, the minimum of $u_{I,>}$ is approached as $r\to\frac{m}{\sqrt{A^2\ell^2-1}}$. Hence 
\begin{equation}
    u_{I,>} > \sqrt{A^2 \ell^2-1} \cot (m\theta)+\frac{A\ell}{\sin (m\theta)}=\cosh \left(\mathrm{sgn}(\cos(m\theta))\cosh^{-1}\left(\frac{1}{\sin{(m\theta)}}\right)+\cosh^{-1}(A\ell)\right)>1.
\end{equation}
Therefore, $L(r,\theta)=B_\alpha \coth^{-1}u_{I,>}$ is real.

For Class II, the argument of $\tanh^{-1}$ is
\begin{equation}    u_{II}=\sqrt{1+A^2\ell^2}\coth(m\theta)+\frac{A \ell^2 m }{r \sqrt{1+A^2 \ell^2}}\frac{1}{\sinh(m\theta)}>\sqrt{1+A^2\ell^2}.
\end{equation}
Hence, $L(r,\theta)=B_\alpha \coth^{-1}u_{II}$ is real.

For Class III, the argument of $\tanh^{-1}$ is
\begin{equation}
    u_{III}=\sqrt{1-A^2 \ell^2} \tanh (m\theta)-\frac{A\ell^2m}{r\sqrt{1-A^2\ell^2}}\frac{1}{\cosh{(m\theta)}}<\sqrt{1-A^2\ell^2}<1.
\end{equation}
Since $0<r<\frac{m}{\sqrt{A^2\ell^2-1}}$ for $f(r)>0$, the minimum of $u_{III}$ is approached as $r\to\frac{m}{\sqrt{A^2\ell^2-1}}$. Hence,
\begin{equation}
    u_{III}>\sqrt{1-A^2 \ell^2} \tanh (m\theta)-\frac{A\ell}{\cosh{(m\theta)}}=\cos\left(\cos^{-1}\left(\tanh(m\theta)\right)+\cos^{-1}\sqrt{1-A^2\ell^2}\right).
\end{equation}

For Class I, at the extremal value of
$X$ given by Eq.~\eqref{rmaxCI} at $\theta=0$, $L=B_\alpha \frac{\pi}{2}$, and $Z$ diverges, which is the boundary of the Poincar\'e patch.


\subsection{Global embeddings of solutions}

We can further consider the following embedding of the C-metric into the covering space, thereby obtaining an embedding for each class that best resembles that of the C-metric. The metric of the covering space is
\begin{equation}
    ds^2 = -dY_{-1}^2-dY_{0}^2+dY_{1}^2 +dY_{2}^2,
\end{equation}
where the constraint
\begin{equation}
    -\ell^2=-Y_{-1}^2-Y_0^2+Y_1^2+Y_2^2
\end{equation}
is imposed on the coordinates $(Y_{-1},X_0,Y_1,Y_2)$. 

AdS spacetime in global coordinates is
\begin{equation}
    ds^2 =- \left(1 + \frac{R^{2}}{\ell^{2}}\right) dT^2 +
    \frac{dR^2}{1 + \frac{R^{2}}{\ell^{2}}}
    +R^2 d\vartheta^2,
\end{equation}
which is obtained by setting
\begin{align}
Y_{-1} &=  \ell\sqrt{1 + \frac{R^{2}}{\ell^{2}}}\cos\!\bigl( T\bigr),\\
Y_{0}  &=  \ell\sqrt{1 + \frac{R^{2}}{\ell^{2}}}\sin\!\bigl( T\bigr),\\
Y_{1}  &=  R \cos\!\bigl(\vartheta\bigr),\\
Y_{2}  &=  R \sin\!\bigl(\vartheta\bigr).
\end{align}
We also have $R=\sqrt{Y_1^2+Y_2^2}$, $\vartheta=\tan^{-1}\left(\frac{Y_2}{Y_1}\right)$, and $T=\tan^{-1}\left(\frac{X_0}{Y_{-1}}\right)$.

For the C-metric Eq.~\eqref{eq:cmetric-polar}, we use coordinates $(\tilde{\tau},\tilde{r},\tilde{\theta})$
and AdS length $\ell$
to distinguish them from the $(\tau,r,\theta)$ coordinates we employ the length scale $\ell_\alpha$ from Eq.~\eqref{Rscal}. We shall only do the embedding for the $x=+1$ identification and not the $x=-1$ identification for both Class I and II as the $x=-1$ identification is completely analogous. 

\paragraph{Class~$\mathrm{I}_\mathrm{slow}$: $A\ell<1$} For Class I, we find
\begin{subequations}
\begin{align}
Y_{-1}  &=\frac{\ell}{\sqrt{1-A^2\ell^2}}\frac{\sqrt{f(\tilde{r})}}{\Omega(\tilde{r},\tilde{\theta})}\cos\left(\tilde{\tau}\sqrt{\frac{1}{A^2\ell^2}-1} \right),\\
Y_0  &= \frac{\ell}{\sqrt{1-A^2\ell^2}}\frac{\sqrt{f(\tilde{r})}}{\Omega(\tilde{r},\tilde{\theta})}\sin\left(\tilde{\tau}\sqrt{\frac{1}{A^2\ell^2}-1}\right),\\
Y_1 &= -\frac{\sqrt{1-A^2 \ell^2} }{A  }+\frac{1}{A \sqrt{1-A^2 \ell^2} \Omega (\tilde{r}\tilde{\theta} )},\\
Y_2  &= \frac{1}{m} \frac{r\sin(m\tilde{\theta})}{\Omega(\tilde{r},\tilde{\theta})}
\end{align}
\end{subequations}
for the C-metric Eq.~\eqref{eq:cmetric-polar}.

For our solution, 
the embedding (up to an $SO(2,2)$ symmetry) is 
\begin{align}
Y_{-1} &= \frac{\ell_\alpha}{2}(H_0+B_\alpha\tanh^{-1}(X(r,\theta))
+\frac{\ell_\alpha}{2(H_0+B_\alpha\tanh^{-1}(X(r,\theta))} \nonumber
\\&\left(1+\cot ^{-1}\left(\sqrt{1-A^2 \ell^2} \cot (m\theta) -\frac{A \ell^2 m}{r \sqrt{1-A^2 \ell^2} }\frac{1}{\sin (m\theta)}\right)^2-\left(\frac{1}{A^2\ell^2}-1\right)\tau^2\right),\\
Y_{0}  &= \frac{\tau \ell_\alpha}{H_0+B_\alpha\tanh^{-1}(X(r,\theta))}\sqrt{\frac{1}{A^2\ell^2}-1},\\
Y_{1}  &= \frac{\ell_\alpha}{2}(H_0+B_\alpha\tanh^{-1}(X(r,\theta))
+\frac{\ell_\alpha}{2(H_0+B_\alpha\tanh^{-1}(X(r,\theta))} \nonumber
\\&\left(-1+\cot ^{-1}\left(\sqrt{1-A^2 \ell^2} \cot (m\theta) -\frac{A \ell^2 m}{r \sqrt{1-A^2 \ell^2} }\frac{1}{\sin (m\theta)}\right)^2-\left(\frac{1}{A^2\ell^2}-1\right)\tau^2\right),\\
Y_{2}  &= \frac{\ell_\alpha}{H_0+B_\alpha\tanh^{-1}(X(r,\theta))} \cot ^{-1}\left(\sqrt{1-A^2 \ell^2} \cot (m\theta) -\frac{A \ell^2 m}{r \sqrt{1-A^2 \ell^2} }\frac{1}{\sin (m\theta)}\right),
\end{align}
where 
\begin{equation}
    X(r,\theta)=(m+A r \cos(m\theta))\left(\frac{r^2}{\ell^2}+\frac{m^2}{1-A^2\ell^2}\right)^{-1/2}.
\end{equation}

\paragraph{Class~$\mathrm{I}_\mathrm{rapid}$: $f(r)>0$, $A\ell>1$} For Class I, we find
\begin{subequations}
\begin{align}
Y_{-1}  &= \frac{\sqrt{A^2 \ell^2-1} }{A  }+\frac{1}{A \sqrt{A^2 \ell^2-1} \Omega (\tilde{r},\tilde{\theta} )},\\
Y_0  &= \frac{\ell}{\sqrt{A^2\ell^2-1}}\frac{\sqrt{f(\tilde{r})}}{\Omega(\tilde{r},\tilde{\theta})}\sinh\left(\tilde{\tau}\sqrt{1-\frac{1}{A^2\ell^2}}\right),\\
Y_1  &= \frac{\ell}{\sqrt{A^2\ell^2-1}}\frac{\sqrt{f(\tilde{r})}}{\Omega(\tilde{r},\tilde{\theta})}\cosh\left(\tilde{\tau}\sqrt{1-\frac{1}{A^2\ell^2}}\right),\\
Y_2  &= \frac{1}{m} \frac{r\sin(m\tilde{\theta})}{\Omega(\tilde{r},\tilde{\theta})}
\end{align}
\end{subequations}
for the C-metric Eq.~\eqref{eq:cmetric-polar}.

For our solution, the embedding (up to an $SO(2,2)$ symmetry) is 
\begin{align}
Y_{-1} &= \frac{\ell_\alpha}{2}(H_0+B_\alpha\tan^{-1}(X(r,\theta))
+\frac{\ell_\alpha}{2(H_0+B_\alpha\tan^{-1}(X(r,\theta))} \nonumber
\\&\left(1+\coth ^{-1}\left(\sqrt{A^2 \ell^2-1} \cot (m\theta) +\frac{A \ell^2 m}{r \sqrt{A^2 \ell^2-1} }\frac{1}{\sin (m\theta)}\right)^2-\left(1-\frac{1}{A^2\ell^2}\right)\tau^2\right),\\
Y_{0}  &= \frac{\tau \ell_\alpha}{H_0+B_\alpha\tan^{-1}(X(r,\theta))}\sqrt{1-\frac{1}{A^2\ell^2}},\\
Y_{1}  &= \frac{\ell_\alpha}{2}(H_0+B_\alpha\tan^{-1}(X(r,\theta))
+\frac{\ell_\alpha}{2(H_0+B_\alpha\tan^{-1}(X(r,\theta))} \nonumber
\\&\left(-1+\coth ^{-1}\left(\sqrt{A^2 \ell^2-1} \cot (m\theta) -\frac{A \ell^2 m}{r \sqrt{1-A^2 \ell^2} }\frac{1}{\sin (m\theta)}\right)^2-\left(1-\frac{1}{A^2\ell^2}\right)\tau^2\right),\\
Y_{2}  &= \frac{\ell_\alpha}{H_0+B_\alpha\tanh^{-1}(X(r,\theta))} \coth ^{-1}\left(\sqrt{A^2 \ell^2-1} \cot (m\theta) +\frac{A \ell^2 m}{r \sqrt{A^2 \ell^2-1} }\frac{1}{\sin (m\theta)}\right).
\end{align}

\paragraph{Class~II: $f(r)>0$} For Class II, we find
\begin{subequations}
\begin{align}
Y_{-1}  &=\frac{\ell}{\sqrt{1+A^2\ell^2}}\frac{\sqrt{f(\tilde{r})}}{\Omega(\tilde{r},\tilde{\theta})}\sinh\left(\tilde{\tau}\sqrt{\frac{1}{A^2\ell^2}+1}\right),\\
Y_0  &=-\frac{\sqrt{1+A^2 \ell^2} }{A  }+\frac{1}{A \sqrt{1+A^2 \ell^2}\Omega(\tilde{r},\tilde{\theta})},\\
Y_1 &=\frac{\ell}{\sqrt{1+A^2\ell^2}}\frac{\sqrt{f(\tilde{r})}}{\Omega(\tilde{r},\tilde{\theta})}\cosh\left(\tilde{\tau}\sqrt{\frac{1}{A^2\ell^2}+1}\right),\\
Y_2  &= \frac{1}{m} \frac{r\sinh(m\tilde{\theta})}{\Omega(\tilde{r},\tilde{\theta})}
\end{align}
\end{subequations}
for the standard C-metric, whereas for our solution, we obtain 
\begin{align}
Y_{-1} &= \frac{\ell_\alpha}{2}(H_0+B_\alpha\tan^{-1}(X(r,\theta))
+\frac{\ell_\alpha}{2(H_0+B_\alpha\tan^{-1}(X(r,\theta))} \nonumber
\\&\left(1+\coth ^{-1}\left(\sqrt{1+A^2 \ell^2} \coth (m\theta) +\frac{A \ell^2 m}{r \sqrt{1+A^2 \ell^2} }\frac{1}{\sinh (m\theta)}\right)^2-\left(\frac{1}{A^2\ell^2}+1\right)\tau^2\right),\\
Y_{0}  &= \frac{\ell_\alpha}{H_0+B_\alpha\tan^{-1}(X(r,\theta))}\sqrt{\frac{1}{A^2\ell^2}+1}\tau,\\
Y_{1}  &= \frac{\ell_\alpha}{2}(H_0+B_\alpha\tan^{-1}(X(r,\theta))
+\frac{\ell_\alpha}{2(H_0+B_\alpha\tan^{-1}(X(r,\theta))} \nonumber
\\&\left(-1+\coth ^{-1}\left(\sqrt{1+A^2 \ell^2} \coth (m\theta) +\frac{A \ell^2 m}{r \sqrt{1+A^2 \ell^2} }\frac{1}{\sinh (m\theta)}\right)^2-\left(\frac{1}{A^2\ell^2}+1\right)\tau^2\right),\\
Y_{2}  &= \frac{\ell_\alpha}{H_0+B_\alpha\tan^{-1}(X(r,\theta))} \cot ^{-1}\left(\sqrt{1+A^2 \ell^2} \coth (m\theta) +\frac{A \ell^2 m}{r \sqrt{1+A^2 \ell^2} }\frac{1}{\sinh (m\theta)}\right),
\end{align}
up to an $SO(2,2)$ symmetry, where 
\begin{equation}
    X(r,\theta)=(m+A r \cosh(m\theta))\left(\frac{r^2}{\ell^2}-\frac{m^2}{1+A^2\ell^2}\right)^{-1/2}.
\end{equation}
 
\paragraph{Class~III: $f(r)>0$}

For the standard C-metric, we find
\begin{subequations}
\begin{align}
Y_{-1}  &=\frac{\ell}{\sqrt{1-A^2\ell^2}}\frac{\sqrt{f(\tilde{r})}}{\Omega(\tilde{r},\tilde{\theta})}\sinh\left(\tilde{\tau}\sqrt{\frac{1}{A^2\ell^2}-1} \right),\\
Y_0  &= \frac{1}{m} \frac{r\cosh(m\tilde{\theta})}{\Omega(\tilde{r},\tilde{\theta})},\\
Y_1 &=\frac{\ell}{\sqrt{1-A^2\ell^2}}\frac{\sqrt{f(\tilde{r})}}{\Omega(\tilde{r},\tilde{\theta})}\cosh\left(\tilde{\tau}\sqrt{\frac{1}{A^2\ell^2}-1}\right),\\
Y_2  &= -\frac{\sqrt{1-A^2\ell^2}}{A}+\frac{1}{A\sqrt{1-A^2\ell^2}\Omega(\tilde{r},\tilde{\theta})},
\end{align}
\end{subequations}
whereas for our solution
\begin{align}
Y_{-1} &= \frac{\ell_\alpha}{2}(H_0+B_\alpha\tan^{-1}(X(r,\theta))
+\frac{\ell_\alpha}{2(H_0+B_\alpha\tan^{-1}(X(r,\theta))} \nonumber
\\&\left(1+\tanh ^{-1}\left(\sqrt{1-A^2 \ell^2} \tanh (m\theta) -\frac{A \ell^2 m}{r \sqrt{1-A^2 \ell^2} }\frac{1}{\cosh (m\theta)}\right)^2-\left(\frac{1}{A^2\ell^2}-1\right)\tau^2\right),\\
Y_{0}  &= \frac{\ell_\alpha}{H_0+B_\alpha\tan^{-1}(X(r,\theta))}\sqrt{\frac{1}{A^2\ell^2}-1}\tau,\\
Y_{1}  &= \frac{\ell_\alpha}{2}(H_0+B_\alpha\tan^{-1}(X(r,\theta))
+\frac{\ell_\alpha}{2(H_0+B_\alpha\tan^{-1}(X(r,\theta))}\nonumber
\\&\left(-1+\tanh ^{-1}\left(\sqrt{1-A^2 \ell^2} \tanh (m\theta) -\frac{A \ell^2 m}{r \sqrt{1-A^2 \ell^2} }\frac{1}{\cosh (m\theta)}\right)^2-\left(\frac{1}{A^2\ell^2}-1\right)\tau^2\right),\\
Y_{2}  &= \frac{\ell_\alpha}{H_0+B_\alpha\tan^{-1}(X(r,\theta))} \tanh ^{-1}\left(\sqrt{1-A^2 \ell^2} \tanh (m\theta) -\frac{A \ell^2 m}{r \sqrt{1-A^2 \ell^2} }\frac{1}{\cosh (m\theta)}\right),
\end{align}
up to an $SO(2,2)$ symmetry, where 
\begin{equation}
    X(r,\theta)=(m+A r \sinh(m\theta))\left(\frac{r^2}{\ell^2}-\frac{m^2}{1+A^2\ell^2}\right)^{-1/2}.
\end{equation}

\section{Proof of the absence of the thin-wall effect in the scalar field equation}

In this appendix, we show a proof that the $\delta (l)$ component of $\mathcal{E}_a^\phi$ is zero. We rewrite the field equation:
\begin{align}
    \mathcal{E}^\phi_a&=G_{ab} \nabla^b\phi  + \nabla_a\phi\left( 
(\nabla\phi)^2 - \Box\phi\right) +\frac{1}{2}
\nabla_a\left((\nabla\phi)^2 \right) \label{Ephiaproof}\\
&=G_{ab} \nabla^b\phi  + \nabla_a\phi\left( 
\nabla_b\phi\nabla^b\phi - \Box\phi\right) +
\nabla_a\nabla_b\phi\nabla^b\phi \\
&=\left(G_{ab}^-\Theta(l)+G^+_{ab}\Theta(-l)+\left(A_{ab}-\frac{1}{2}Ag_{ab}\right)\delta(l) \right)(\Theta(l)\nabla^b\phi^-+\Theta(-l)\nabla^b\phi^+) 
\nonumber\\&+ (\Theta(l)\nabla_a\phi^-+\Theta(-l)\nabla_a\phi^+)\Big( 
(\Theta(l)\nabla_b\phi^-+\Theta(-l)\nabla_b\phi^+)(\Theta(l)\nabla^b\phi^-+\Theta(-l)\nabla^b\phi^+) 
\nonumber\\&- (\Box\phi^-\Theta(l)+\Box\phi^+\Theta(-l)+n^\theta[\nabla_\theta\phi]\delta(l))\Big) +
(\nabla_a\nabla_b\phi^+\Theta(-l)+\nabla_a\nabla_b\phi^-\Theta(l)\nonumber\\
&+n_a[\nabla_b\phi])(\Theta(l)\nabla^b\phi^-+\Theta(-l)\nabla^b\phi^+).
\end{align}
Taking advantage of the fact that 
\begin{eqnarray}
    \Theta(\pm l)^2=\Theta(\pm l),\;\;\Theta(l)\Theta(-l)=0,
\end{eqnarray}
we obtain  
\begin{align}
    \mathcal{E}^\phi_a&=\underbrace{\left(G^-_{ab} \nabla^b\phi^-  + \nabla_a\phi^-\left( 
(\nabla\phi^-)^2 - \Box\phi^-\right) +\frac{1}{2}
\nabla_a\left((\nabla\phi^-)^2 \right) \right)\Theta(l)}_{\mathcal{E}_a^{-\phi}}\nonumber\\
&+\underbrace{\left(G^+_{ab} \nabla^b\phi^+  + \nabla_a\phi^+\left( 
(\nabla\phi^+)^2 - \Box\phi^+\right) +\frac{1}{2}
\nabla_a\left((\nabla\phi^+)^2 \right) \right)\Theta(-l)}_{\mathcal{E}_a^{+\phi}}\nonumber\\
&+\delta (l)(...).
\end{align}
It is implicit that we have been solving for $\mathcal{E}_a^{-\phi}$, which is zero. Due to the mirror symmetry, $\mathcal{E}_a^{+\phi}$ is also zero.

If we include the boundary within the domain, $\mathcal{E}^\phi_a$ may be decomposed into
\begin{equation}    \mathcal{E}^\phi_a=\mathcal{E}^\phi_a{}^{(0)}+\mathcal{E}^\phi_a{}^{(1)}\delta(l).
\end{equation}
$\mathcal{E}^\phi_a{}^{(0)}$, as stated, is zero. Then, using Eqs.~\eqref{eq47} and \eqref{eq48}, we have
\begin{equation}
    \frac{1}{2}\nabla_a((\nabla\phi)^2)=\nabla_a\nabla_b\phi\nabla^b\phi=\text{bounded terms }+n_a[\nabla_b\phi]\nabla^b\phi\cdot\delta(l),
\end{equation}
where "bounded" refers to terms that are finite and piecewise smooth as $l \to 0$. Such terms remain $\mathcal{O}(1)$ at the wall and therefore do not generate any $\delta(l)$ contribution. 
\begin{align}
    \mathcal{E}^\phi_a{}^{(1)}&= G_{ab}\nabla^b\phi-\nabla_a\phi\,\Box \phi+\frac{1}{2}\nabla_a((\nabla\phi)^2) \text{ terms of order $\delta(l)$}\\
    &=(A_{ab}-\frac{1}{2}A g_{ab})\nabla^b\phi \cdot\delta(l)-\nabla_a\phi (n^\theta[\nabla_\theta\phi])\delta(l)+n_a[\nabla_b\phi]\nabla^b\phi\cdot\delta(l)\\
    &=(A_{ab}-\frac{1}{2}A g_{ab}-n^\theta[\nabla_\theta\phi] g_{ab}+n_a[\nabla_b\phi])\nabla^b\phi \cdot\delta(l).\label{eq62}
\end{align}

Here, note that 
\begin{equation}
    \kappa_{ab} \equiv [g_{ab,c}]n^c=[g_{ab,\theta}]n^\theta.
\end{equation}
We have previously calculated for the equation of state that $A_{ab}$ is diagonal and
\begin{align}
    A_{\tau\tau}-\frac{1}{2}Ag_{\tau\tau}&=\frac{1}{2}\kappa^r{}_{r}g_{\tau\tau},\\
    A_{rr}-\frac{1}{2}Ag_{rr}&=\frac{1}{2}\kappa^{\tau}{}_{\tau}g_{rr},\\
    A_{\theta\theta}-\frac{1}{2}Ag_{\theta\theta}&=0.
\end{align}

Keeping in mind that all the relevant tensors are diagonal, we have
\begin{align}    \mathcal{E}^\phi_\tau{}^{(1)}=\left(\frac{1}{2}\kappa^r{}_rg_{\tau\tau}-n^\theta[\nabla_\theta\phi] g_{\tau\tau}\right)\nabla^\tau\phi\cdot\delta(l)=0,
\end{align}
since $g_{\tau \tau}\nabla^\tau\phi=\nabla_\tau\phi=0$.

For the angular term
\begin{equation}
    \mathcal{E}^\phi_\theta{}^{(1)}=(0-n^\theta[\nabla_\theta\phi] g_{\theta\theta}+n_\theta[\nabla_\theta\phi])\nabla^\theta\phi \cdot\delta(l)=0.
\end{equation}
In addition, we have 
\begin{equation}
    \phi=\frac{1}{2}\ln g_{\tau\tau}+\text{const.}
\end{equation}
Hence, it is straightforward to show that 
\begin{equation}
    n^\theta[\nabla_\theta\phi] =\frac{1}{2g_{\tau\tau}}\kappa_{\tau\tau}=\frac{1}{2}\kappa^\tau{}_\tau.\label{eq65}
\end{equation}
Inserting Eq.~\eqref{eq65} into Eq.~\eqref{eq62}, we obtain  
\begin{align}
    \mathcal{E}^\phi_r{}^{(1)}&=\left(\frac{1}{2}\kappa^\tau{}_\tau g_{rr}-\frac{1}{2} g_{rr}\kappa^\tau{}_\tau+0\right)\nabla^r\phi \cdot\delta(l)=0,
\end{align}
and hence, $\mathcal{E}^{\phi(1)}_a$ is zero.

\newpage

\bibliographystyle{unsrt}
\bibliography{refs}

\end{document}